\documentclass[aps,pra%,preprint%
,groupedaddress]{revtex4-1}

\usepackage{amssymb,amsmath}
\usepackage{natbib}
\usepackage[usenames,dvipsnames]{xcolor}
\usepackage{graphicx}
\usepackage[utf8]{inputenc}

\newcommand{\potential}{\sigma(\vec{x})}
\newcommand{\fieldPhi}{\hat{\Phi}(\vec{x},t)}
\newcommand{\fieldPhiPr}{\hat{\Phi}(\vec{x}^{\,\prime},t)}
\newcommand{\shortPhi}{\hat{\Phi}}
\newcommand{\shortPhiPr}{\hat{\Phi}^{\,\prime}}
\newcommand{\fieldpsi}{\psi_j(\vec{x})}
\newcommand{\fieldpsistar}{\psi_j^{*}(\vec{x}^{\,\prime})}
\newcommand{\EMTOp}{\hat{T}_{\mu\nu}(\vec{x},t)}
\newcommand{\TnnOp}{\hat{T}_{00}(\vec{x},t)}
\newcommand{\Tnn}{T_{00}(\vec{x},t)}
\newcommand{\TnnCP}{T_{00}(\vec{x}_{cp},t)}
\newcommand{\TzzOp}{\hat{T}_{zz}(\vec{x},t)}
\newcommand{\Tzz}{T_{zz}(\vec{x},t)}
\newcommand{\TzzCP}{T_{zz}(\vec{x}_{cp},t)}
\newcommand{\pslim}{\lim_{\vec{x}^{\,\prime}\to\vec{x}}}
\newcommand{\deltalim}{\lim_{\vec{\Delta}\to0}}
\newcommand{\psnabla}{\vec{\nabla}\cdot\left(\vec{\nabla}+\vec{\nabla}^{\prime}\right)}
\newcommand{\cpnabla}{\vec{\nabla}_{cp}^2}
\newcommand{\momint}{\int\frac{\mathrm{d}^{d}p}{\left(2\pi\right)^{d}}}
\newcommand{\Gsigma}{G_{\sigma}(\vec{x},\vec{x}^{\,\prime}, k)}
\newcommand{\Gzero}{G_{0}(\vec{x},\vec{x}^{\,\prime}, k)}
\newcommand{\Geff}{G(\vec{x},\vec{x}^{\,\prime}, k)}

\newcommand{\EQref}[1]{Eq.~\eqref{#1}}
\newcommand{\tabref}[1]{Tab.~\ref{#1}}
\newcommand{\figref}[1]{Fig.~\ref{#1}}
\newcommand{\secref}[1]{Sec.~\ref{#1}}
\DeclareMathOperator{\ImP}{Im}
\DeclareMathOperator{\ReP}{Re}
\DeclareMathOperator{\Erf}{Erf}
\DeclareMathOperator{\er}{e}
\DeclareMathOperator{\Max}{max}
\DeclareMathOperator{\Sign}{sgn}

\begin{document}

\title{Worldline Numerics for Energy-Momentum Tensors in Casimir Geometries}

\author{Marco Sch{\"a}fer}
\email[]{marco.jschaefer@gmail.com}
\affiliation{Theoretisch-Physikalisches Institut, Abbe Center of Photonics, \\ Friedrich-Schiller-Universit\"at Jena, Max-Wien-Platz 1, 07743 Jena, Germany}
%\thanks{}
\author{Idrish Huet}
\email[]{idrish@ifm.umich.mx}
%\thanks{}
\affiliation{Facultad de Ciencias en F\'{\i}sica y Matem\'aticas, Universidad Aut\'onoma de Chiapas, \\
Ciudad Universitaria, Tuxtla Guti\'errez 29050, M\'exico}
\affiliation{Theoretisch-Physikalisches Institut, Abbe Center of Photonics, \\ Friedrich-Schiller-Universit\"at Jena, Max-Wien-Platz 1, 07743 Jena, Germany}
\author{Holger Gies}
\email[]{holger.gies@uni-jena.de}
\affiliation{Theoretisch-Physikalisches Institut, Abbe Center of Photonics, \\ Friedrich-Schiller-Universit\"at Jena, Max-Wien-Platz 1, 07743 Jena, Germany}
\affiliation{Helmholtz-Institut Jena, Fr\"obelstieg 3, 07743 Jena, Germany}
%\thanks{}

\date{\today}

\begin{abstract}
We develop the worldline formalism for computations of composite
operators such as the fluctuation induced energy-momentum tensor. As
an example, we use a fluctuating real scalar field subject to
Dirichlet boundary conditions. The resulting worldline representation
can be evaluated by worldline Monte-Carlo methods in continuous
spacetime. We benchmark this worldline numerical algorithm with the
aid of analytically accessible single-plate and parallel-plate Casimir
configurations, providing a detailed analysis of statistical and
systematic errors. The method generalizes straightforwardly to
arbitrary Casimir geometries and general background potentials.
\end{abstract}
%\pacs{}
%\keywords{}

\maketitle

%%% Main Matter=================================================================

\section{Introduction\label{sec:intro}}

  The worldline formalism
  \cite{Feynman:1950ir,Halpern:1977ru,Bern:1990ux,Strassler:1992zr} is
  a field theory technique with various computational advantages such
  as the reduction of the number of diagrams in perturbative
  expansions. It is particularly powerful for amplitude computations
  in external backgrounds
  \cite{Schmidt:1993rk,Reuter:1996zm,Shaisultanov:1995tm,Schubert:2001he},
  as string-inspired techniques become highly efficient in this
  case. Moreover, fluctuation-induced quantities such as quantum
  actions, energies, and forces can also be efficiently computed by
  means of Monte Carlo simulations of the worldline path
  integral. Even fully nonperturbative problems can be tackled within
  the worldline formalism
  \cite{Affleck:1981bma,Nieuwenhuis:1996mc,Gies:2005sb,Dietrich:2013kza,Bastianelli:2014bfa,Dietrich:2015oba}.

  The numerical method has given unprecedented access to
  background problems with nontrivial spacetime dependencies in QED
  \cite{Gies:2001zp,Gies:2001tj,Langfeld:2002vy,Schmidt:2003bf,Mazur:2014hta}, QCD \cite{Deschner:2008,Epelbaum:2015cca}
  fermionic systems \cite{Dunne:2009zz}, as well as the physics of
  Casimir forces
  \cite{Gies:2003cv,Gies:2006bt,Gies:2006xe,Gies:2006cq,Weber:2009dp,Schaden:2009zza,Schaden:2009zz,Aehlig:2011xg};
  analogous analytical results for QED have also come within reach
  using worldline instanton approximations \cite{Dunne:2005sx,Dunne:2006st,Dietrich:2007vw,Strobel:2014tha,Linder:2015vta,Ilderton:2015qda,Dumlu:2015paa}. Important
  advantages of the Monte-Carlo method are that spacetime is not
  discretized but remains continuous in the algorithm, and
  renormalization can be performed on the level of finite and
  numerically controllable quantities. In addition the worldline
  formalism offers an intuitive picture of quantum fluctuations, as
  the worldlines can be viewed as spacetime trajectories of the
  fluctuating particles. The numerical method has also been applied to
  Minkowski-valued 2-point functions for external legs \cite{Gies:2011he}.

  In the present work, we demonstrate that the technique can also
  be extended to composite operators, using the induced
  energy-momentum tensor (EMT) as an example. The quest for such a
  method was first initiated in \cite{QFEXT11}. The new difficulty 
  arises from the fact that standard tools such as point-splitting
  lead to expressions involving the 2-point function of the
  fluctuating field (internal line). As shown below, this can
  nevertheless be dealt with in a comparatively straightforward
  fashion by using open in addition to closed worldlines.

  The present conceptual and in large parts technical work is
  motivated by the computation of energy-momentum tensors for Casimir
  configurations. This problem is paradigmatic as the EMT sources the
  gravitational field. Since induced Casimir energies are typically
  negative (if associated with an attractive force)
  \cite{Casimir,Bordag20011,Milton:2001yy,Milton:2008st,Bordag:2009zzd},
  information about the EMTs for Casimir configurations is a testbed
  for conceptual considerations of the interplay of quantum field
  theory and gravity, see, e.g.,
  \cite{Fulling:2007xa,Milton:2007ar,Shajesh:2007sc} for a discussion
  of how the Casimir energy ``falls''.

  Generically, many fundamental investigations of dynamical
  spacetime properties start with assumptions imposed on the
  properties of the EMT on the right-hand side of Einstein's
  equations. Most prominently, in order to exclude exotic phenomena
  \cite{Morris1988,PhysRevLett.61.1446,PhysRevD.39.3182,WormholeVisser,PhysRevD.72.044023,PhysRevLett.81.3567},
  energy conditions (ECs) on the properties of the EMT are
  imposed. For instance, certain ``mechanisms'' for superluminal
  travel can be excluded if an energy condition of the form
  $T_{\mu\nu}V^{\mu}V^{\nu}\geq0$ is imposed, where $V^{\mu}$ is the
  tangent vector of a geodesic $\gamma$. If $V^{\mu}$ is a null vector
  this condition is called the null energy condition (NEC). The
  Casimir effect violates some energy conditions such as the NEC
  \cite{Graham:2002yr,Olum:2002ra,Graham:2005cq,Graham:2006mx}. It can
  therefore not be used to rule out superluminal travel. In fact, superluminal (but still causality
  preserving) phase velocities are known to occur in the
  parallel-plate Casimir configuration
  \cite{Scharnhorst:1990sr,Barton:1989dq,Barton:1992pq,Dittrich:1998fy,Dittrich:2000zu}. 
  The Casimir configurations studied so far at least obey the
  weaker averaged NEC (ANEC). For a further discussion of this topic,
  see
  \cite{Graham:2005cq,Graham:2007va,Visser:1996iw,Visser:1996iv,Visser:1996ix,Visser:1997sd,Fewster:2006uf,Graham:2007va}. In
  the present work, we use the violation of the NEC by the Casimir
  effect as an illustration for our new computational method. For
  conventional computational strategies to determine the induced EMT,
  for example, by means of mode summation or expansion, image charge methods,
  or similar techniques, see, e.g.,
  \cite{Brown:1969na,Milton:2002vm,Milton:2004vp,Scardicchio:2005di}.

This paper is intended to be a manual for the numerical computation of
composite operators using the worldline formalism. It is organized as
follows: in \secref{sec:comp_op} we apply the worldline formalism to
composite operators, specifically the EMT of a real scalar field
obeying Dirichlet boundary conditions (BCs). We discuss the
differences between the formalism for composite and non-composite
operators or functionals like the effective action. In
\secref{sec:SglPl} we test our numerical worldline algorithm by
calculating components of the energy-momentum tensor for a single
plate with Dirichlet BCs analytically and numerically. The numerical
calculation of the EMT and of the NEC is presented for the parallel
plate configuration in \secref{sec:CasPl}. We compare our numerical
results with known analytical results and discuss the arising
systematic and statistical errors. We conclude with a summary of our
results and an outlook on future worldline calculations.

\section{Worldline Formalism for the energy-momentum tensor\label{sec:comp_op}}

Composite, or local, operators in quantum field theory are local products of field operators and their derivatives. Being distribution-valued, their product at the same spacetime point may be ill-defined and give rise to divergences. Such operators can nevertheless be used for calculations after regularizing the divergences, for example, by point-splitting, $\zeta$ function, or dimensional regularization. In the following, we compute the vacuum expectation value of the energy-momentum tensor operator of a scalar field; a composite operator constructed from the scalar field operator and its derivatives.

Our presentation of the basic formalism follows the one of \cite{Scardicchio:2005di} with only a few differences. We study a quantum scalar field $\fieldPhi$ that is a $\mathcal{C}^{\infty}$ map from a domain $\mathcal{D}$ of $d+1$-dimensional Minkowski spacetime $\mathcal{M}^{(d,1)}$ onto the linear space of self-adjoint operators on the Fock space $\mathcal{F}$
\begin{align}
\fieldPhi:\ \mathcal{M}^{(d,1)}\ \ni\ \mathcal{D}\quad\rightarrow\quad\mathcal{F}.
\end{align}
$\fieldPhi$ has mass $m$ and is minimally coupled to a static classical background potential $\potential$. Later on $\potential$ will impose boundary conditions on the fluctuations of $\fieldPhi$. We use the $d+1$-dimensional Minkowski metric $g_{\mu\nu}$ with the "mostly minus" signature $\left(+,-,\ldots,-\right)$. From the Lagrangian density operator
\begin{align}\label{eq:Lagrangian}
\hat{\mathcal{L}} = {} & \frac{1}{2} \partial_\mu\fieldPhi\partial^\mu\fieldPhi-\frac{1}{2}\left(m^2+\potential\right)\fieldPhi\fieldPhi,
\end{align}
we derive the equation of motion for $\fieldPhi$,
\begin{align}\label{eq:PhiEoM}
\left(\partial_\lambda\partial^\lambda+m^2+\potential\right)\fieldPhi = {} & 0,
\end{align}
as well as its canonical energy-momentum tensor operator ($\shortPhi:=\fieldPhi$)
\begin{align}
\begin{split}
\EMTOp = {} & 
\frac{\partial\hat{\mathcal{L}}}{\partial(\partial^\mu\shortPhi)}\partial_\nu\shortPhi-g_{\mu\nu}\hat{\mathcal{L}}\label{eq:EMTOpDef}\\
= {} & \partial_\mu\shortPhi\partial_\nu\shortPhi-\frac{1}{2}g_{\mu\nu}\left(\partial^\lambda\shortPhi\partial_\lambda\shortPhi-(m^2+\potential)\shortPhi\shortPhi\right).
\end{split}
\end{align}
The EMT operator in \EQref{eq:EMTOpDef} contains products of field operators at the same spacetime point that can lead to divergences. We regularize them by a point-splitting procedure in the spatial components $\vec{x}$, i.e., 
\begin{align}
\fieldPhi\fieldPhi  \longrightarrow  \fieldPhi\fieldPhiPr &\quad&
\partial_\alpha\fieldPhi\partial_\beta\fieldPhi  \longrightarrow  \partial_\alpha\fieldPhi\partial_\beta^{\,\prime}\fieldPhiPr.
\end{align}
The point-split EMT operator is then ($\shortPhiPr:=\fieldPhiPr$)
\begin{align}
\EMTOp = {} & \label{eq:EMT1}
\pslim\left[\partial_\mu\shortPhi\partial^{\,\prime}_\nu\shortPhiPr-\frac{1}{2}g_{\mu\nu}\left(\partial^\lambda\partial^{\,\prime}_\lambda-m^2-\potential\right)\shortPhi\shortPhiPr\right].
\end{align}

Our final goal is to compute the effects of boundary conditions on the energy-momentum tensor. For the sake of convenience, we restrict ourselves to computing those EMT components that contain information relevant for the null energy condition (NEC) in the $z$ direction, where the $z$ coordinate is always the $d$th coordinate of our spatial vectors $\vec{x}=(x^{1},\ldots,x^{d})=(x^{1},\ldots,z)$. This null energy condition is given by the vacuum expectation value of the projection of the EMT onto a null curve with the null tangent vector $V^{\mu}=(1,0,\ldots,0,1)$, 
\begin{align}\label{eq:NEC}
0\leq{}\left\langle\EMTOp V^{\mu}V^{\nu}\right\rangle = {} & \left\langle\TnnOp+\TzzOp\right\rangle{}={}\Tnn+\Tzz,
\end{align}
where $\left\langle\cdots\right\rangle$ denotes the vacuum expectation value. All our calculations generalize straightforwardly to other EMT components.
In order to compute the vacuum expectation values of \EQref{eq:EMT1}, we expand $\fieldPhi$ in terms of spatial eigenmodes and corresponding
creation and annihilation operators $\hat{a}_j^\dagger$ and
$\hat{a}_j$, 
\begin{align}
\fieldPhi = {} & \sum_j 
 \frac{1}{\sqrt{2E_j}}\left(\fieldpsi\er^{iE_j t}\hat{a}_j + \psi^*_j(\vec{x})\er^{-iE_j t}\hat{a}_j^\dagger \right),
\end{align}
where a discrete notation is used for both discrete and continuous parts of the spectrum for simplicity.
The $\fieldpsi$ are defined as normalized eigenmodes of the Laplacian in the presence of the potential $\potential$
\begin{align}\label{eq:Helmholtz}
\left(-\vec{\nabla}^2
+m^2 + \potential\right)\fieldpsi = {} & E_j^2 \fieldpsi .
\end{align}
It is convenient to parametrize the energy eigenvalues $E_j^2$
  also in terms of momenta $k_j$ according to $E_j^2=k_j^2+m^2$. Then, the eigenvalue equation reads $(-\vec{\nabla}^2+\potential)\fieldpsi=k_j^2\fieldpsi$, which is a Helmholtz-type equation.
The corresponding Green's function equation,
\begin{align}\label{eq:DefGsigma}
\left(-\vec{\nabla}^2+\potential-k^2\right)\Gsigma = {} & \delta\left(\vec{x}-\vec{x}^{\,\prime}\right),
\end{align}
is solved by the spectral representation
\begin{align}\label{eq:SpecGsigma}
\Gsigma = {} & \sum_j 
\frac{\fieldpsi\fieldpsistar}{k_j^2-k^2-i\varepsilon}.
\end{align}
We aim at expressing \EQref{eq:NEC} in terms of $\Gsigma$. We therefore insert unity in the form
\begin{align}
1 = \int\limits_0^\infty\mathrm{d}k\,2k\,\delta\left(k_j^{2}
-k^2\right) 
= {} & \int\limits_0^\infty\mathrm{d}k\,2k\,\lim_{\varepsilon\to0^+}\frac{1}{\pi}
\ImP\left(\frac{1}{k_j^2-k^2-i\varepsilon}\right),
\end{align}
and exchange the $k$ integration and the $j$ summation. 
This yields
\begin{align}\label{eq:EMTGsigma}
\begin{split}
\Tnn\big\vert_{\sigma} = {} & \pslim\int\limits_0^\infty\mathrm{d}k\,\frac{k}{\pi}\left(E_k+\frac{1}{2E_k}\psnabla\right)
\ImP \sum_j 
\frac{\fieldpsi\fieldpsistar}{k_j^2-k^2-i\varepsilon}\\
\Tzz\big\vert_{\sigma} = {} & \pslim\int\limits_0^\infty\mathrm{d}k\,\frac{k}{\pi}\left(\frac{1}{E_k}\partial_z\partial_{z^\prime}-\frac{1}{2E_k}\psnabla\right)\ImP \sum_j 
\frac{\fieldpsi\fieldpsistar}{k_j^2-k^2-i\varepsilon}.
\end{split}
\end{align}
The sum inside the imaginary part is just $\Gsigma$. In general, instead of unity a decaying exponential $\er^{-k/\Lambda}$ must be inserted in order to construct local counterterms for renormalization (cf. \cite{Scardicchio:2005di}). $\Lambda$ then serves as a cutoff for large momenta $k$. This is, however, not necessary in our calculations because we are going to evaluate the EMT only at points for which $\potential=0$, so that all local counterterms, that is, all counterterms that depend on $\potential$, are automatically zero. Furthermore, since the term $\fieldpsi\fieldpsistar$ is in general complex, pulling it into the argument of the imaginary part generates an additional term that is proportional to $\ImP\fieldpsi\fieldpsistar$. This term vanishes in the limit $\vec{x}\to\vec{x}^{\,\prime}$ and hence will be ignored.
The effects of boundary conditions imposed on the field fluctuations by $\potential$ are described by the difference between the EMT with non-vanishing potential and the EMT with $\potential=0$. So far, we have left the potential $\potential$ arbitrary to emphasize that our calculations are independent of its specific properties. We can therefore repeat all the above steps with a vanishing potential. The only changes that occur are the mode functions $\fieldpsi$ in \EQref{eq:Helmholtz}. The corresponding Green's function is $\Gzero$, defined by \EQref{eq:DefGsigma} for $\potential=0$, rather than $\Gsigma$. We switch to the \textit{common point} variables 
\begin{align}\label{eq:CPvariables}
\vec{x}_{cp}:=\frac{\vec{x}+\vec{x}^{\,\prime}}{2}, {}\qquad & {}\qquad \vec{\Delta}:=\frac{\vec{x}-\vec{x}^{\,\prime}}{2},
\end{align}
and with $\Geff=\Gsigma-\Gzero$, evaluated at $\vec{x}=\vec{x}_{cp}+\vec{\Delta}$ and $\vec{x}^{\,\prime}=\vec{x}_{cp}-\vec{\Delta}$ we have
\begin{align}
\begin{split}
\TnnCP = {} & \deltalim\int\limits_0^\infty\mathrm{d}k\,\frac{k}{\pi}\left(E_k+\frac{1}{4E_k}\cpnabla\right)\ImP\Geff,\\
\TzzCP = {} & \deltalim\int\limits_0^\infty\mathrm{d}k\,\frac{k}{\pi}\left(\frac{1}{4E_k}\left(\partial_{z_{cp}}^2-\partial_{\Delta_z}^2\right)-\frac{1}{4E_k}\cpnabla\right)\ImP\Geff\label{eq:EMTG}.
\end{split}
\end{align}
Equation \eqref{eq:EMTG} is independent of any specific mode expansion of $\fieldPhi$. It only depends on the Green's function $\Geff$, which is representation-independent by definition. Therefore, any method for computing $\Geff$ can be used at this stage, e.g., an optical approach in \cite{Scardicchio:2005di}. The subtraction of $\Gzero$ also removes the divergent terms that are independent of the potential $\potential$ and normalizes the EMT such that it vanishes for $\potential=0$.

\subsection{The worldline representation of $\Geff$\label{sec:WL_for_G}}

\begin{figure}[ht]
\centerline{\includegraphics[width=0.95\textwidth]{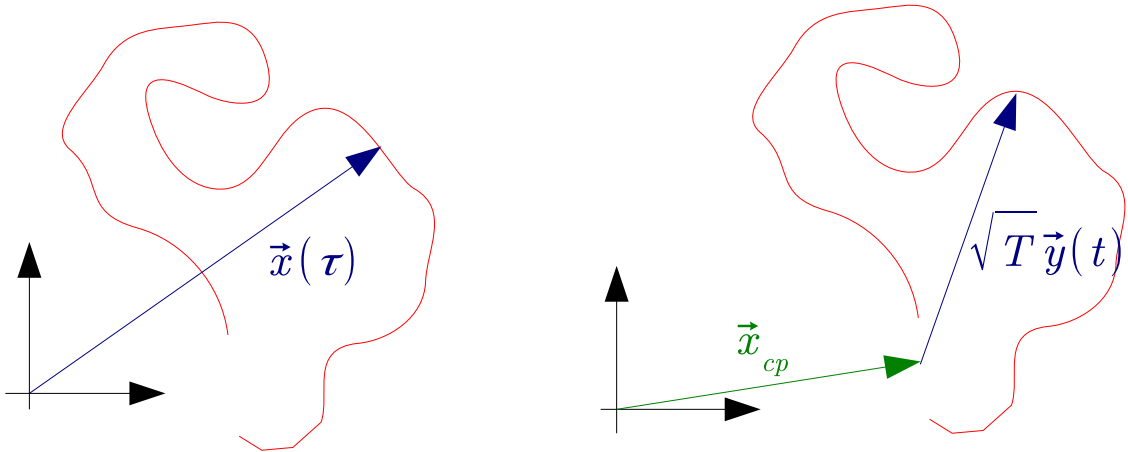}}
\caption[Rescaling the worldlines to unit loops]{Schematic depiction of the rescaling from the worldlines $\vec{x}(\tau)$ to the paths $\vec{x}_{cp}+\sqrt{T}\vec{y}(t)$ with $\tau=T\,t$ and the unit worldlines $\vec{y}(t)$. \label{Pic:UnitLoops}}
\end{figure}

For simplicity, we consider here only the massless case $m=0$. Our calculations can straightforwardly be carried over to the massive case, for details, see App.~\ref{sec:App1}. In order to express the Green's function in the worldline formalism, we interpret $\Geff$ as the matrix element of an operator $\hat{\mathcal{G}}(k)$ and \EQref{eq:Helmholtz} as a quantum mechanical Schr{\"o}dinger problem whose Hamiltonian is $H=-\vec{\nabla}^2+\potential$. Then $\Geff$ corresponds to a quantum mechanical propagator, Fourier transformed to energy space, from which the free motion has been subtracted. Hence, it can be written in position space in terms of a propertime representation:
\begin{align}
\Geff 
= {} & \left\langle\vec{x}^{\,\prime}\right\vert\hat{\mathcal{G}}(k)\left\vert\vec{x}\right\rangle
%\nonumber \\
= {}%
% & 
 i\int\limits_0^\infty\mathrm{d}s\,\er^{isk^2}\left\langle\vec{x}^{\,\prime}\right\vert\er^{-is\left(-\vec{\nabla}^2+\potential\right)}-\er^{is\vec{\nabla}^2}\left\vert\vec{x}\right\rangle. \label{eq:TransAmp}
\end{align}
The matrix element in \EQref{eq:TransAmp} is the quantum mechanical
transition amplitude of a fictitious particle moving from $\vec{x}$ at
the fictitious time $\tau=0$ to $\vec{x}^{\,\prime}$ at $\tau=s$ with
a Hamiltonian $H=-\vec{\nabla}^2+\potential$. The corresponding
Feynman path integral in position space is then straightforward to
find. Next, we perform formal Wick rotations in both the
$s$ and $k$ planes, such that $s=-iT$ and $k=-ik_{E}$, which are
consistent with causality. This casts $\Geff$ into its doubly Wick-rotated form
\begin{align}
 G (\vec{x}, \vec{x}^{\,\prime}, -i k_E) = {} &
\label{eq:GworldlineB}
\int\limits_0^\infty\mathrm{d}T\,\er^{-Tk_E^2}\int\limits_{\vec{x}=\vec{x}(0)}^{\vec{x}^{\,\prime}=\vec{x}(T)}\mathcal{D}{\vec{x}}(\tau)\er^{-\int\limits_0^T\mathrm{d}\tau\frac{\dot{\vec{x}}^2}{4}}
\left(\er^{-\int\limits_0^T\mathrm{d}\tau\,\sigma(\vec{x}(\tau))}-1\right).
\end{align}
 The variables $s$ and $T$ are called Minkowskian and Euclidean propertime, respectively. Both describe the time evolution of the fictitious Schr{\"o}dinger problem, but neither is a physical, measurable time. An analogous propertime representation for the effective action has been used in previous calculations of effective interaction energies for the Casimir effect and similar boundary configurations \cite{Gies:2005sb,Gies:2006cq,Gies:2006xe}. The worldline representation of the effective action contains, however, a path integral over closed loops, whereas for $\Geff$, open worldlines running from $\vec{x}$ to $\vec{x}^{\,\prime}$ must be computed. The implicitly normalized path integral in \EQref{eq:GworldlineB} gives, for the free path integral, the standard free propagator,
\begin{align}\label{eq:freePI}
\int\limits_{\vec{x}=\vec{x}(0)}^{\vec{x}^{\,\prime}=\vec{x}(T)}\mathcal{D}{\vec{x}}(\tau)\er^{-\int\limits_0^T\mathrm{d}\tau\frac{\dot{\vec{x}}^2}{4}} = {} &
\frac{\er^{-\vec{\Delta}^2/T}}{\left(4\pi T\right)^{d/2}} \qquad\text{with}\quad \vec{\Delta}=\frac{\vec{x}-\vec{x}^{\,\prime}}{2}.
\end{align}
From this normalization one derives the shorthand notation for the path integral expectation value, the worldline average of an arbitrary operator $\mathcal{O}\left[\vec{x}(\tau)\right]$:
\begin{align}\label{eq:PIvev}
\left<\mathcal{O}\left[\vec{x}(\tau)\right]\right>_{\vec{x},\vec{x}^\prime}:= {} & \int\limits_{\vec{x}=\vec{x}(0)}^{\vec{x}^\prime=\vec{x}(T)}\mathcal{D}{\vec{x}}(\tau)\er^{-\int\limits_0^T\mathrm{d}\tau\frac{\dot{\vec{x}}^2}{4}} \mathcal{O}\left[\vec{x}(\tau)\right]\ \biggr{/} \ \frac{\er^{-\vec{\Delta}^2/T}}{\left(4\pi T\right)^{d/2}}.
\end{align}
Pulling out the imaginary part prescription in front of the $k$ integral in \EQref{eq:EMTG}, we can perform both Wick rotations for the components of the EMT. Details can be found in App.~\ref{sec:App1}. We finally arrive at the manifestly real worldline
expressions for the normalized vacuum expectation values of $\TnnOp$
and $\TzzOp$,
\begin{align}\label{eq:EMTcomp}
\begin{split}
\TnnCP = {} &
\frac{1}{(4\pi)^{\frac{d+1}{2}}}\deltalim\int\limits_0^\infty\frac{\mathrm{d}T}{4T^{\frac{d+1}{2}}}\left(-\frac{2}{T}+\cpnabla\right)\er^{-\frac{\vec{\Delta}^2}{T}}\Big\langle\er^{-\int\limits_0^T\mathrm{d}\tau\,\sigma(\vec{x}(\tau))}-1\Big\rangle_{\vec{x}_{cp},\vec{\Delta}}\\
\TzzCP = {} &
\frac{1}{(4\pi)^{\frac{d+1}{2}}}\deltalim\int\limits_0^\infty\frac{\mathrm{d}T}{4T^{\frac{d+1}{2}}}
\left(\partial_{z_{cp}}^2-\partial_{\Delta_z}^2-\cpnabla\right)\er^{-\frac{\vec{\Delta}^2}{T}}\Big\langle\er^{-\int\limits_0^T\mathrm{d}\tau\,\sigma(\vec{x}(\tau))}-1\Big\rangle_{\vec{x}_{cp},\vec{\Delta}}.
\end{split}
\end{align}
The expectation value of the path integral in \EQref{eq:EMTcomp} is defined by \EQref{eq:PIvev} with the operator \begin{math}\mathcal{O}\left[\vec{x}(\tau)\right] = \mathrm{exp}\left(-\int_0^T\mathrm{d}\tau\,\sigma(\vec{x}(\tau))\right)-1\end{math}. It is an average over worldlines $\vec{x}(\tau)$ that start at $\vec{x}=\vec{x}(0)$ and end at $\vec{x}^{\,\prime}=\vec{x}(T)$, and that are weighted with a Gau{\ss}ian velocity distribution. It is convenient to switch to coordinates such that for $\vec{x}\to\vec{x}^{\,\prime}$ all paths are fixed with respect to one common point $\vec{x}_{cp}=\left(\vec{x}+\vec{x}^{\,\prime}\right)/2$ (cf. Fig.~\ref{Pic:UnitLoops}). We therefore call these worldlines \textit{common point} loops or lines. The remaining path is written in terms of a dimensionless unit worldline $\vec{y}(t)$ with $t\in\left[0,1\right]$ and $\tau=T\,t$. 
The unit line $\vec{y}(t)$ itself can be written as the sum of the
classical path from $\vec{x}$ to $\vec{x}^{\,\prime}$ and a deviation $\vec{\mathcal{Z}}(t)$ that obeys
$\vec{\mathcal{Z}}(0)=0=\vec{\mathcal{Z}}(1)$. We choose the origin of
the $\vec{y}$ coordinate system to be the point $\vec{x}_{cp}$,
\begin{align}\label{eq:UnitLoops}
\vec{x}(\tau) = \vec{x}(Tt) = {} & \vec{x}_{cp}+\sqrt{T}\vec{y}(t) = {} 
\vec{x}_{cp}+\sqrt{T}\left((1-2t)\vec{\delta}+\vec{\mathcal{Z}}(t)\right).
\end{align}
In the limit $\vec{\delta}\to0$ the worldline $\vec{y}(t)$ closes, it becomes a loop $\vec{y}(t)\to\vec{\mathcal{Z}}(t)$. We have set above
\begin{align}
\frac{\vec{\Delta}}{\sqrt{T}} = \frac{\vec{x}-\vec{x}^{\,\prime}}{2\sqrt{T}} = {} & \vec{\delta}.
\end{align}
The $\sqrt{T}$ dependence in this relation is crucial for the computation of the $\Delta_z$-derivatives in \EQref{eq:EMTcomp}. Rescaling the worldlines, $\vec{x}(\tau) \to \vec{y}(t)$, makes the weight factor of the resulting $\vec{y}$ path integral independent of $T$. This allows us to compute the
expectation value by generating only one ensemble of unit worldlines
$\vec{y}(t)$, which are all defined with respect to the same
coordinate system. The worldline average is now calculated by
replacing the path integral with a sum over a finite number $N$ of
unit worldlines that are themselves approximated by a finite number
$n_{ppl}$ of points per line $\vec{y}_i$ with
$i\in\{0,\ldots,n_{ppl}\}$. The points $\vec{y}_i$ are random numbers
distributed according to the Gau{\ss}ian weight
factor \begin{math}\mathrm{exp}(-\int_0^1\mathrm{d}t\,\dot{\vec{y}}^{\,2}/4)\end{math},
  being a discretized realization of an open line with the vector $2\vec{\delta}$
  connecting its endpoints. As a result, the expectation value of an
operator $\mathcal{O}$ is written as an average,
\begin{align}\label{eq:PIAverage}
\Big\langle\mathcal{O}\Big\rangle_{\vec{x}_{cp},\vec{\Delta}} = {} & \frac{1}{N}\sum\limits_{l=1}^{N}\mathcal{O}_{l}(\vec{y}_i),\qquad i\in\{1,\ldots,n_{ppl}\}.
\end{align} 
The common point paths $\vec{y}_i$ are conveniently generated with the
$d$ loop algorithm \cite{Gies:2005sb}, which also works for open
worldlines.

\subsection{Dirichlet constraints on $\partial\mathcal{D}$\label{sec:Dirichlet_constraint}}

The entire formalism that we have outlined so far works, of course, for arbitrary static background field configurations $\potential$. In our calculations, we use
\begin{align}\label{eq:potential}
\potential = {} & \lambda\int\limits_{\partial\mathcal{D}}\mathrm{d}\Sigma\,\delta\left({\vec{x}-\vec{x}_{\partial\mathcal{D}}}\right),
 \quad {\vec{x}_{\partial\mathcal{D}}\in \partial\mathcal{D}},
\end{align}
where $\mathrm{d}\Sigma$ is a surface element on $\partial\mathcal{D}$. We recover Dirichlet BCs in the limit $\lambda\to\infty$ (cf. \cite{Graham:2003ib,Graham:2002fw}). The subtraction of the vacuum Green's function in \EQref{eq:EMTG} already removed all divergences independent of $\potential$. Therefore, $\Geff$ can only diverge at points for which $\potential\neq0$ and the energy-momentum tensor in \EQref{eq:EMTcomp} is finite on $\mathcal{D}$, where $\sigma$ vanishes. All remaining divergences are located on the boundary $\partial\mathcal{D}$ and can be related to the infinite amount of energy necessary to constrain $\fieldPhi$ on all momentum scales to fulfill the Dirichlet boundary condition on $\partial\mathcal{D}$.

The parametrization \EQref{eq:potential} and the subsequent Dirichlet limit greatly simplify the worldline average because now we have
\begin{align}
\begin{split}
 \Big\langle\er^{-T\int\limits_0^1\mathrm{d}t\,\sigma(\vec{x}_{cp}+\sqrt{T}\vec{y}(t))}-1\Big\rangle_{\vec{x}_{cp},\vec{\Delta}} = {} &
\left\{\begin{array}{cl} -1 & \mbox{if $\vec{x}_{cp}+\sqrt{T}\vec{y}(t)$ intersects $\partial\mathcal{D}$}\\ 0 & \mbox{{otherwise}} \end{array}\right\}\\
= {} &
-\Big\langle\Theta\left[\,\mathcal{S}\,(\vec{x}_{cp}+\sqrt{T}\vec{y}(t))\,\right]\Big\rangle_{\vec{x}_{cp},\vec{\Delta}}.
\end{split}\label{eq:ThetaIntersect}
\end{align}
Equation (\ref{eq:ThetaIntersect}) states that only paths
$\vec{x}_{cp}+\sqrt{T}\vec{y}(t)$ which violate the boundary
conditions lead to deviations from the trivial vacuum and thus
contribute to the expectation value. We call the function
$\mathcal{S}$ the intersection condition. It gives a geometric
description of how and for which values of $T$ the worldlines
intersect the boundary. As long as the start and end point of a given
worldline are not on the boundary
($\vec{x}_{cp}\pm\vec{\Delta}\notin\partial\mathcal{D}$), the
intersection condition will always determine a minimal non-zero value
for the propertime $T_{min}>0$ for which the worldline first intersects
$\partial\mathcal{D}$. $T_{min}$ denotes the minimum propertime that
is necessary for a worldline to propagate from the start to the end
point and intersect a boundary in between. For any $T<T_{min}$
deviations from the straight line between $\vec{x}$ and
$\vec{x}^{\,\prime}$ are typically strongly suppressed. Only for
sufficiently large propertimes $T\geq T_{min}$ does the diffusive
Brownian motion process, described by the path integral, create
sufficiently large random detours that can intersect
$\partial\mathcal{D}$. In the propertime integral, $T_{min}$ serves as
a lower bound and removes the divergence for $T\to0$. From a physics
point of view, $T_{min}$ acts as an ultraviolet cutoff because small
propertimes correspond to large momenta. The intersection
  condition $\mathcal{S}$ may also provide an upper bound $T_{max}$. This value $T_{max}$ would then mark 
the propertime for which the worldlines
no longer intersect the boundary. This needs to be considered
especially for configurations where the boundary consists of compact objects 
(see e.g. \cite{Gies:2006cq}).

\subsection{Compact expressions of $\TnnCP$ and $\TzzCP$ for worldline numerics\label{sec:CompExpr}}

The EMT components are completely finite on $\mathcal{D}\backslash\partial\mathcal{D}$ by virtue of \EQref{eq:ThetaIntersect}. We may therefore decompose $\TnnCP$ and $\TzzCP$ further and study the resulting, more compact, terms one by one.
Toward this end, we parametrize the intersection condition as \begin{math}\mathcal{S}\,(\vec{x}_{cp}+\sqrt{T}\vec{y}(t))\,=\,\sqrt{T}\mathcal{M}-1\end{math}, where the function $\mathcal{M}$ has inverse length dimension and describes the geometrical conditions for a worldline to intersect the boundary $\partial\mathcal{D}$. It depends on $\vec{x}_{cp}$ and $\vec{\delta}$, and through the latter also on $T$. This $T$ dependence is not explicitly known but it vanishes as $\vec{\Delta}\to0$. In general, $\mathcal{M}$ is not a smooth function of these parameters. In fact, it can be non-differentiable in either variable. Since $\mathcal{M}$ depends on the shape of the boundary $\partial\mathcal{D}$ for the setup under consideration, its functional properties must be investigated each time anew. In addition, there can be several ways in which the intersection condition $\mathcal{S}$ may be written. Whereas different parametrizations are equivalent, they can exhibit very different properties during numerical evaluation. Thus, the following calculations should be understood as primarily formal. First, we assume the function $\mathcal{M}$ to be differentiable twice in both $\vec{x}_{cp}$ and $\vec{\delta}$. We additionally assume that all limits can be evaluated, and that $\mathcal{M}$ only determines a lower bound $T_{min}$ on the propertime integral. In this case, we have \begin{math} T_{min} = \mathcal{M}^{-2}\end{math}.

We now insert \EQref{eq:ThetaIntersect} into Eqs.~(\ref{eq:EMTcomp}) and find
\begin{align}
\begin{split}\label{eq:EMTcompTheta}
\TnnCP = {} &
-\frac{1}{4}\int\limits_0^\infty\frac{\mathrm{d}T}{(4\pi T)^{\frac{d+1}{2}}}\left(-\frac{2}{T}+\cpnabla\right)\er^{-\frac{\vec{\Delta}^2}{T}}\Big\langle\Theta\left(\,\sqrt{T}\mathcal{M}-1\,\right)\Big\rangle_{\vec{x}_{cp}}\bigg\vert_{\vec{\Delta}\to0},\\
\TzzCP = {} &
-\frac{1}{4}\int\limits_0^\infty\frac{\mathrm{d}T}{(4\pi T)^{\frac{d+1}{2}}}
\left(\partial_{z_{cp}}^2-\partial_{\Delta_z}^2-\cpnabla\right)\er^{-\frac{\vec{\Delta}^2}{T}}\Big\langle\Theta\left(\,\sqrt{T}\mathcal{M}-1\,\right)\Big\rangle_{\vec{x}_{cp}}\bigg\vert_{\vec{\Delta}\to0}.
\end{split}
\end{align}
In order to evaluate the propertime integral, we need to interchange several limits, which requires justification. First, we note that all expressions we deal with are finite by construction due to the Dirichlet constraint \EQref{eq:ThetaIntersect}. Since this is valid for all $\vec{\Delta}$, the limit $\vec{\Delta}\to0$ may be interchanged with all other limits except $\partial_{\Delta_z}$. Furthermore, since we approximate the worldline average by a finite sum, this average can be exchanged with other limits and be conveniently computed last. We may also formally interchange propertime integration and differentiations because $T$, $\vec{x}_{cp}$ and $\vec{\Delta}$ are independent variables. There is, however, the intersection condition $\mathcal{S}=\sqrt{T}\mathcal{M}-1$, which depends on all three variables. The function $\mathcal{M}$ also depends on the combination $\vec{\delta}=\vec{\Delta}/\sqrt{T}$. As a consequence, the derivatives of $\mathcal{S}$, or more specifically of $\mathcal{M}$, must be computed before the propertime integration (cf. \secref{sec:App2}). Picturing the worldlines as paths in space helps examine the situation: we need not only compute the intersection condition itself but also its derivatives, that is, we need to determine how the intersection is altered if $\vec{x}_{cp}$ or $\vec{\Delta}$ are changed. A derivative with respect to $\vec{x}_{cp}$ can be viewed geometrically as moving the complete worldline through space without changing its shape. By contrast, the $\vec{\Delta}$ derivative corresponds to opening and closing the worldline at a fixed point $\vec{x}_{cp}$ in space, changing its shape in the process. The required order in which these manipulations of the worldline expressions should be performed is then:
\begin{enumerate}
\item{compute the derivatives of $\Theta\left(\,\sqrt{T}\mathcal{M}-1\right)$ with respect to $z_{cp}$ and $\delta_z=\Delta_z/\sqrt{T}$,}
\item{let $\vec{\Delta}\to0$, that is, $\vec{\delta}\to0$,}
\item{perform the propertime integration, and}
\item{average the expression over all worldlines in the ensemble.}
\end{enumerate}
With these consideration at hand, we write $\TnnCP=\TnnCP\big\vert_{I}+\TnnCP\big\vert_{II}$ where we define
\begin{subequations}
\begin{align}\label{eq:T00I}
\TnnCP\big\vert_{I}\  := {} &
\frac{1}{(4\pi)^{\frac{d+1}{2}}}\frac{1}{d+1}\left\langle\mathcal{M}^{d+1}\right\rangle_{\vec{x}_{cp},\vec{\Delta}=0},\\
\label{eq:T00II}
\TnnCP\big\vert_{II} := {} &
-\frac{1}{2}\,\frac{1}{(4\pi)^{\frac{d+1}{2}}}\frac{1}{d-1}\left\langle\vec{\nabla}^2_{cp}\mathcal{M}^{d-1}\right\rangle_{\vec{x}_{cp},\vec{\Delta}=0}.
\end{align}
\end{subequations}
With $\TzzCP$ we proceed accordingly and write it as a sum of four terms: 
\begin{subequations}
\begin{align}\label{eq:TzzIa}
\TzzCP\big\vert_{Ia} := {} &
-\frac{1}{2}\,\frac{1}{(4\pi)^{\frac{d+1}{2}}}\frac{1}{d-1}\left\langle\partial^2_{z_{cp}}\mathcal{M}^{d-1}\right\rangle_{\vec{x}_{cp},\vec{\Delta}=0}\\
\TzzCP\big\vert_{Ib} := {} &
\frac{1}{2}\,\frac{1}{(4\pi)^{\frac{d+1}{2}}}\frac{1}{d+1}\left\langle\partial_{\delta_{z}}^2 \mathcal{M}^{d+1}\right\rangle_{\vec{x}_{cp},\vec{\Delta}=0},\label{eq:TzzIb}\\
\TzzCP\big\vert_{Ic} := {} &
-\frac{1}{(4\pi)^{\frac{d+1}{2}}}\frac{1}{d+1}\left\langle\mathcal{M}^{d+1}\right\rangle_{\vec{x}_{cp},\vec{\Delta}=0}\,=\, - \TnnCP\big\vert_{I}\, ,\label{eq:TzzIc}\\
\label{eq:TzzII}
\TzzCP\big\vert_{II} := {} &
\frac{1}{2}\,\frac{1}{(4\pi)^{\frac{d+1}{2}}}\frac{1}{d-1}\left\langle\vec{\nabla}^2_{cp}\mathcal{M}^{d-1}\right\rangle_{\vec{x}_{cp},\vec{\Delta}=0}\,=\, -\TnnCP\big\vert_{II}\, .
\end{align}
\end{subequations}
The fourth term, $\TzzCP\big\vert_{Ic}$, comes from acting with $\partial^2_{\Delta_z}$ on the exponential $\er^{-\vec{\Delta}^2/T}$. Another term, the mixed term $(\partial_{\Delta_z}\er^{-\vec{\Delta}^2/T})\partial_{\Delta_z}\langle\cdots\rangle$ vanishes as $\vec{\Delta}\to0$.

We note that, with the help of the above decomposition into compact terms, the null energy condition along the $z$ axis \EQref{eq:NEC} now reduces to computing
\begin{align}\label{eq:GeneralNEC}
0\leq{}\Tnn+\Tzz = {} & \TzzCP\big\vert_{Ia} + \TzzCP\big\vert_{Ib}.
\end{align}

\section{The energy-momentum tensor for a single plate\label{sec:SglPl}}

The numerical computation of $\Tnn$ and $\Tzz$ in $d=2$ and $d=3$ space dimensions in the case that $\partial\mathcal{D}$ is a single $d-1$-dimensional surface, i.e., a plate, is our first proof-of-principle example. The plate imposes Dirichlet BCs on the fluctuations of $\fieldPhi$. It is placed at $z=0$ such that its normal is the $z$ axis. The single Dirichlet plate configuration is also sometimes referred to as the perfect mirror. 

\subsection{Analytic calculation for a single plate\label{sec:AnaSinglePlate}}

Before we use worldline numerics, we compute the EMT for the single Dirichlet plate analytically. For that, we use \EQref{eq:EMTG} and compute $\Geff$ by solving the equation of motion \EQref{eq:Helmholtz} for different boundary conditions. Denoting the BCs by a superscript $\sigma$, we must solve
\begin{align}\label{eq:EoManalytical}
\left(-\vec{\nabla}^{\,2}-p^2\right)\psi_p^\sigma(\vec{x})&{} = 0.
\end{align}
Toward that end, we decompose any $\vec{x}\in\mathcal{D}$ in the $d-1$-dimensional vector $\vec{x}_{||}$ parallel to $\partial\mathcal{D}$ and the $z$ component of $\vec{x}$. The solution of \EQref{eq:EoManalytical} in the half space $z>0$ with Dirichlet boundary conditions at $z=0$ is then
\begin{align}
\psi_p^\sigma(\vec{x})
{} & = \frac{1}{i\sqrt{2}}\left(\exp\left(i\vec{p}\cdot\vec{x}\right)-\exp\left(i\vec{p}\cdot\widetilde{\vec{x}}\right)\right)
\end{align}
where $\vec{x}=\left(\vec{x}_{||},z\right)$ and $\widetilde{\vec{x}}=\left(\vec{x}_{||},-z\right)$. We assumed that there are only outgoing waves at spatial infinity (Sommerfeld radiation condition). In the same manner, the free solution without boundary conditions is found to be \begin{math} \psi_p^{0}(\vec{x}) {}  = \exp\left(i\vec{p}\cdot\vec{x}\right) \end{math}. Both solutions are normalized for $z>0$ and $z^{\prime}>0$, \begin{math} \momint\,\psi_p^\sigma(\vec{x})\,\psi_p^{\sigma\,*}(\vec{x}^{\,\prime}) {}  = {}\delta^d\left(\vec{x}-\vec{x}^{\,\prime}\right) {} = \momint\,\psi_p^0(\vec{x})\,\psi_p^{0\,*}(\vec{x}^{\,\prime}) \end{math}. The Green's functions $\Gsigma$ and $\Gzero$ are now computed according to the spectral representation \EQref{eq:SpecGsigma}. We perform the momentum integration in polar coordinates and find for $\Geff=\Gsigma-\Gzero$ (see also \cite{Sommerfeld,Jackson, HandbuchLinDGL})
\begin{align}
\begin{split}
\Geff\big\vert^{d=2} = {} &
\frac{1}{2\pi}\left(\mathcal{K}_0(-ik\vert\vec{x}-\vec{x}^{\,\prime}\vert)-\mathcal{K}_0(-ik\vert\vec{x}-\widetilde{\vec{x}}^{\,\prime}\vert)\right)\ -\ \frac{1}{2\pi}\mathcal{K}_0(-ik\vert\vec{x}-\vec{x}^{\,\prime}\vert),\\
\Geff\big\vert^{d=3} = {} &
\frac{1}{4\pi}\left(\frac{\er^{ik\vert\vec{x}-\vec{x}^{\,\prime}\vert}}{\vert\vec{x}-\vec{x}^{\,\prime}\vert}-\frac{\er^{ik\vert\vec{x}-\widetilde{\vec{x}}^{\,\prime}\vert}}{\vert\vec{x}-\widetilde{\vec{x}}^{\,\prime}\vert}\right)\ -\ \frac{1}{4\pi}\frac{\er^{ik\vert\vec{x}-\vec{x}^{\,\prime}\vert}}{\vert\vec{x}-\vec{x}^{\,\prime}\vert}.
\end{split}
\end{align}
The $\mathcal{K}_0$ are modified Bessel functions of the second kind (Macdonald functions).

We can now solve \EQref{eq:EMTG} analytically and even use the decomposition of $\TnnCP$ and $\TzzCP$ that we developed in \secref{sec:CompExpr}. Denoting the distance from the plate with $z_{cp}$, we find for the EMT in $d=2$
\begin{subequations}
\begin{align}
\TnnCP\big\vert_{I}^{d=2} & = \hspace{0.8em}\frac{1}{32\pi}\frac{1}{z_{cp}^{\ 3}}, &
\TnnCP\big\vert_{II}^{d=2}\hspace{0.6em} & =  -\frac{1}{16\pi}\frac{1}{z_{cp}^{\ 3}},\\
\TzzCP\big\vert_{Ia}^{d=2} & =  -\frac{1}{16\pi}\frac{1}{z_{cp}^{\ 3}}, &
\TzzCP\big\vert_{Ib+Ic}^{d=2} & = {}  \quad 0.
\end{align}
\end{subequations}
And in the case of 3 spatial dimensions the values for $\TnnCP$ and $\TzzCP$ are
\begin{subequations}
\begin{align}
\TnnCP\big\vert_{I}^{d=3} & = \hspace{0.8em}\frac{1}{32\pi^2}\frac{1}{z_{cp}^{\ 4}}, & 
\TnnCP\big\vert_{II}^{d=3}\hspace{0.6em} & = -\frac{3}{32\pi^2}\frac{1}{z_{cp}^{\ 4}},\\
\TzzCP\big\vert_{Ia}^{d=3} & =  -\frac{3}{32\pi^2}\frac{1}{z_{cp}^{\ 4}}, &
\TzzCP\big\vert_{Ib+Ic}^{d=3} & = \quad 0.
\end{align}
\end{subequations}
We note that $\TzzCP\big\vert_{Ib}$ and $\TzzCP\big\vert_{Ic}$ cannot be calculated separately in a direct manner since we used the functional structure of the worldline representation of $\Geff$ to define these functions. Despite that, we can always compute $\TzzCP\big\vert_{Ib}$ from the sum $\TzzCP\big\vert_{Ib+Ic}$ using the fact that $\TzzCP\big\vert_{Ic}=-\TnnCP\big\vert_{I}$. We thus have 
\begin{align}
\TzzCP\big\vert_{Ib}^{d=2} & = \hspace{0.8em}\frac{1}{32\pi}\frac{1}{z_{cp}^{\ 3}}, &
\TzzCP\big\vert_{Ib}^{d=3} & = \hspace{0.8em}\frac{1}{32\pi^2}\frac{1}{z_{cp}^{\ 4}}.
\end{align}
According to \EQref{eq:GeneralNEC} the NEC along the $z$ axis is then violated,
\begin{align}
\TnnCP+\TzzCP {} = & \left\{\renewcommand{\arraystretch}{1.0}\begin{array}{cc} \mbox{{$-1/\left(32\pi\,z_{cp}^{\,3}\right)$}} & \mbox{\ for $d=2$}\\ \mbox{{$-1/\left(16\pi^2\,z_{cp}^{\,4}\right)$}} & \mbox{\ for $d=3$}\end{array}\renewcommand{\arraystretch}{1}\right. .
\end{align}
These are the same values for the NEC which were derived in \cite{Graham:2005cq}.

\subsection{Worldline calculation for one Dirichlet plate\label{sec:WLSinglePlate}}

\begin{figure}[!ht]
\centerline{\includegraphics[width=0.95\textwidth]{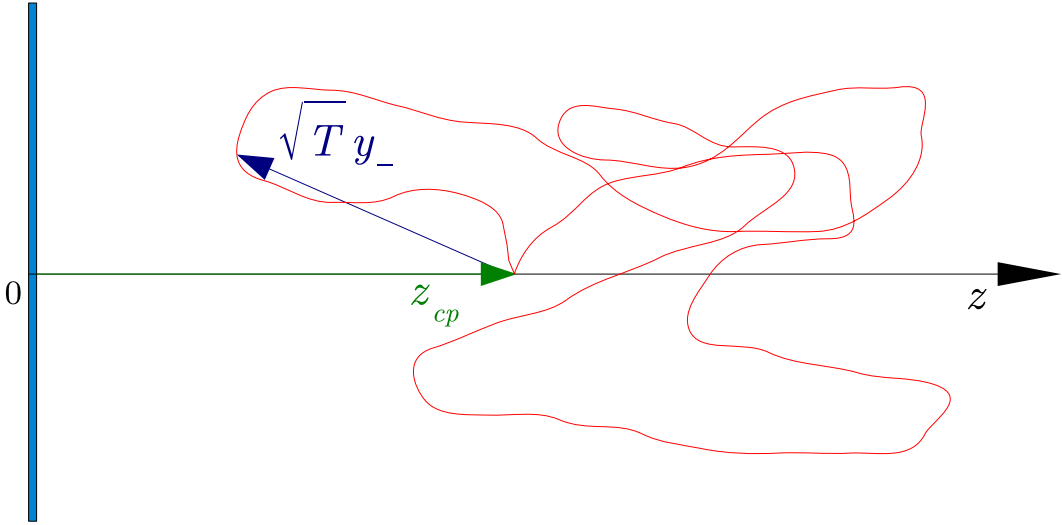}}
\caption[Single plate setup]{Sketch of the single plate setup with an exemplary unit worldline.\label{pic:SinglePlate}}
\end{figure}

The first step in all our worldline calculations is determining the intersection condition $\mathcal{S}=\sqrt{T}\mathcal{M}-1$. For the single plate setup, this is easily done from \figref{pic:SinglePlate}.
The worldlines $\vec{y}(t)$ start at the point $z_{cp}$ and intersect the plate at $z=0$ for all $T$ that fulfill $\sqrt{T}y_{-}+z_{cp}\leq0$. We call the $z$ component of the point on the loop that is closest to the plate $y_{-}$. It is negative for our choice of coordinates in the setup \figref{pic:SinglePlate}. Thus we find for the $\Theta$ function
\begin{align}
\sqrt{T}y_{-}+z_{cp}\leq0 {} \quad\Longrightarrow\quad & \Theta\left(\sqrt{T}\mathcal{M}-1\right)
{} = {}\Theta\left(\sqrt{T}\left(-y_{-}\right)/z_{cp}-1\right).
\end{align}
This corresponds to a minimal propertime
$T_{min}=z_{cp}^2/(-y_{-})^2$, where $(-y_{-})$ is the positive
distance that measures the extension of the loop towards the
plate. $T_{min}$ only depends on the $z$ component of $\vec{y}(t)$
because the plate constrains the propagation of $\vec{y}(t)$ only in
the $z$ direction. For this reason, and because the worldline
distributions factorize with respect to their position space
components, we only need to calculate 1-dimensional
loops. Furthermore, only one point of every loop, the point $y_{-}$,
needs to be found. In this case, we have $T_{max}\to\infty$, i.e., 
if the intersection condition is fulfilled for $T_{min}$, it will be
fulfilled for all $T>T_{min}$.

The 1-dimensional open line $y(t)$ depends on $\vec{\delta}$ and thus on the ratio $\Delta_z/\sqrt{T}$.
Being an extremal point on this loop, $y_{-}:=y(t_{-})$ carries the same dependence. The defining equation for the extremum $y_{-}$ is (cf. \EQref{eq:UnitLoops})
\begin{align}\label{eq:YminDef}
y_{-}:=y(t=t_{-},\delta_z)\qquad\Longrightarrow\qquad\ \frac{\mathrm{d}y}{\mathrm{d}t}\Big\vert_{t_{-}} = {} &
\partial_t\mathcal{Z}(t)\big\vert_{t_{-}}-2\delta_z = 0.
\end{align}
This is an implicit equation for $t_{-}$ which is itself a function of $\delta_z$. Consequently, the minimal point depends on $\delta_z$ in an explicit and implicit way, that is \begin{math} y_{-} = {}  y(t_{-}(\delta_z), \delta_z)\end{math}. In order to compute $\TnnCP$ and $\TzzCP$, we insert $\mathcal{M}=(-y_{-})/z_{cp}$ into \EQref{eq:T00I}-\eqref{eq:T00II} and \EQref{eq:TzzIa}-\eqref{eq:TzzII}. In $\TnnCP\big\vert_{I}$, only a power of $\mathcal{M}$ must be computed, which can straightforwardly be implemented numerically. There are also no difficulties in calculating $\TnnCP\big\vert_{II}$ and $\TzzCP\big\vert_{Ia}$ as $\mathcal{M}$ is obviously differentiable in $z_{cp}$. 

However, $\mathcal{M}$ is strictly speaking not differentiable with respect to $\delta_z$ because the Brownian bridge $y(t)$ is not differentiable. This problem is caused by a naive interchange of derivatives and the path integral. Our strategy is to cure such problems by identifying a constructive definition of the derivatives in terms of geometric properties of the discretized worldlines. For the moment, let us write
\begin{align}
\TzzCP\big\vert_{Ib} {} = & \frac{1}{(4\pi)^{\frac{d+1}{2}}}\frac{\left\langle(-y_{-})^{d}(-y_{-})^{\prime\prime}+d(-y_{-})^{d-1}(-y_{-})^{\prime}(-y_{-})^{\prime}\right\rangle}{2\,z_{cp}^{d+1}}
\end{align}
with $(\cdot)^{\prime}=\partial_{\delta_z}(\cdot)$.
We compute the derivatives with the help of \EQref{eq:UnitLoops} and \eqref{eq:YminDef} and find
\begin{align}
\label{eq:YminVelo}
(-y_{-})^{\prime} := {} &
\frac{\mathrm{d}\ \ }{\mathrm{d}\delta_z}(-y_{-})\Big\vert_{\delta_z=0}\,=\ -\left(1-2t_{-}\right)\Big\vert_{\delta_z=0},\\%[1em]
\label{eq:YminAccel}
(-y_{-})^{\prime\prime} := {} &
\frac{\mathrm{d}^2\ }{\mathrm{d}\delta^2_z}(-y_{-})\Big\vert_{\delta_z=0}\,=\ 2\frac{t_{-}(\delta_z+h_\delta)-t_{-}(\delta_z-h_\delta)}{2h_\delta}\bigg\vert_{\delta_z=0,h_\delta\to0}.
\end{align}
The right-hand sides of these equations are, in fact,
  well-behaved, such that they can be used as a definition of the
  derivatives of $\vec{y}(t)$ to be inserted into the path integral.
The constraint \EQref{eq:YminDef} allows for an analytical form of
$(-y_{-})^{\prime}$. For its computation only the time parameter
$t_{-}(\delta_z=0)$ must be determined. A similar closed form of
$(-y_{-})^{\prime\prime}$ does not exist because the function
$t_{-}(\delta_z)$ is unknown. It can only be calculated numerically by
using the difference quotient in \EQref{eq:YminAccel} with
$\delta_z=0$ and sufficiently small $h_\delta$.  The
$\partial_{\delta_z}$ derivative terms are not necessary for the
computation of the NEC in the case of the single plate because we know
$\TzzCP\big\vert_{Ic}=-\TnnCP\big\vert_{I}$ (in general) and from the
analytical results $\TzzCP\big\vert_{Ib+Ic}=0$ (for the single
plate). Hence, we have $\TzzCP\big\vert_{Ib}=\TnnCP\big\vert_{I}$ and
we know that $\TnnCP\big\vert_{I}$ contains no derivatives. We use
this as an additional check for computing derivatives with our
algorithms. The null energy condition along the $z$ axis can
consequently be read off from \begin{math}\TnnCP+\TzzCP = {}
  \TzzCP\big\vert_{Ia} + \TnnCP\big\vert_{I}\end{math}.

\subsection{Numerical results for $\TnnCP$ and $\TzzCP$}

We need to compute six different worldline averages, $\left\langle(-y_{-})^p\right\rangle$ with $p\in\left\{1,2,3,4\right\}$ and $\left\langle\partial^2_{\delta_z}(-y_{-})^q\right\rangle$ with $q\in\left\{3,4\right\}$. As with any Monte-Carlo method there are statistical and systematic errors. The statistical error arises from the finite number of worldlines $N$ in the ensemble. We use the standard deviation of the observables as a measure for the statistical error, which decreases as $1/\sqrt{N}$. Since the computational costs of our worldline algorithms scale linearly with $N$, the statistical error is readily controlled.

\begin{figure}[!ht]
\includegraphics[width=0.48\textwidth]{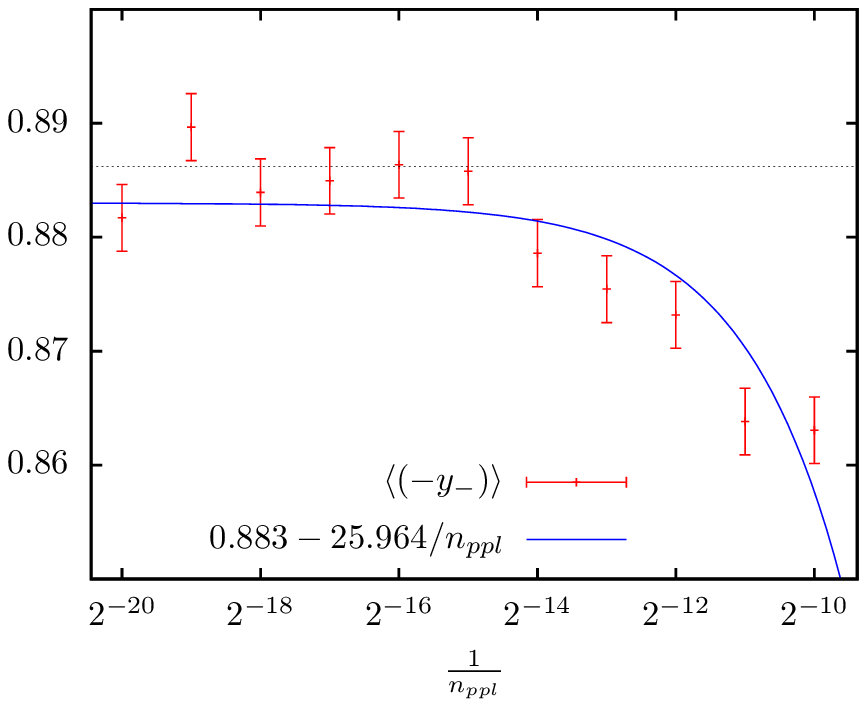}\hspace{12pt}\includegraphics[width=0.48\textwidth]{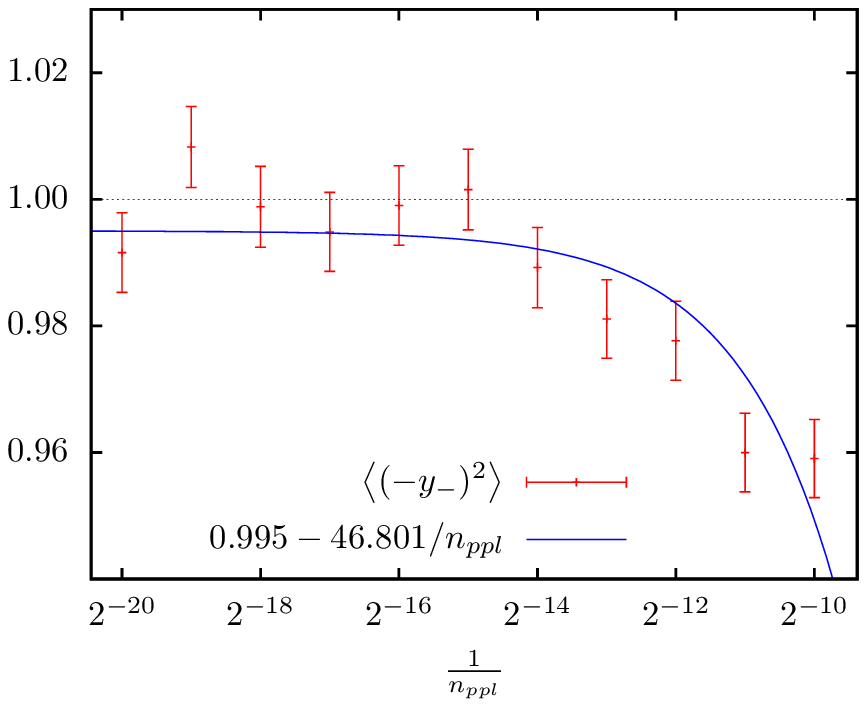}
\newline
\includegraphics[width=0.48\textwidth]{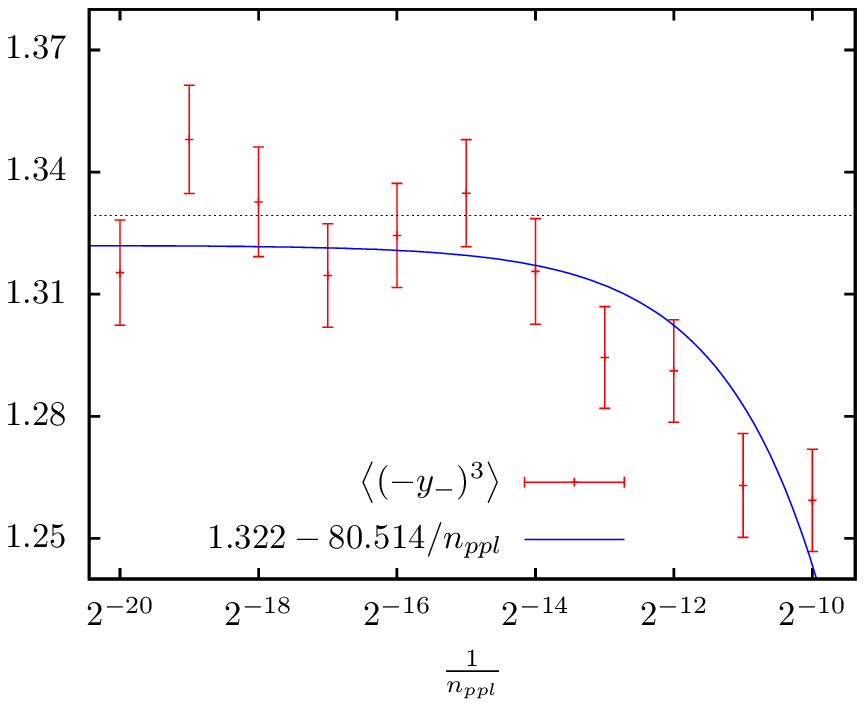}\hspace{12pt}\includegraphics[width=0.48\textwidth]{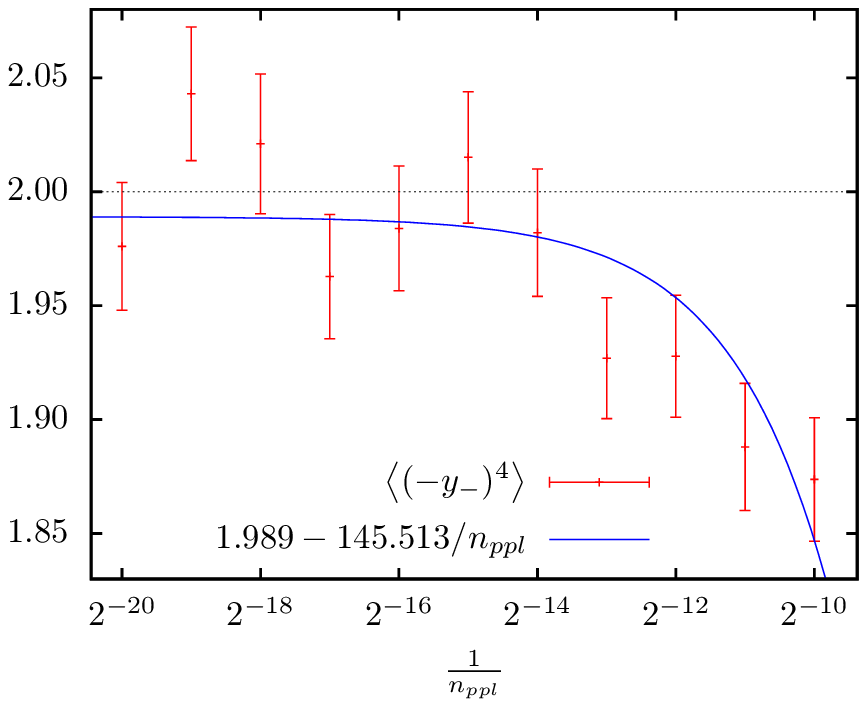}
\caption[Systematic errors of cumulants of $(-y_{-})$]{Numerical
  values for $\left\langle(-y_{-})^r\right\rangle$ with
  $r\in\{1,2,3,4\}$ for $n_{ppl}=2^{10}\ldots 2^{20}$ with error
    bars showing their statistical errors. The solid curves are fits
  to our data and the dotted lines are the respective analytical
  values. Systematic errors are the difference between analytic
  results and our fits for $n_{ppl}\to0$.\label{pic:MP1Error}}
\end{figure}

\begin{figure}[!ht]
\includegraphics[width=0.48\textwidth]{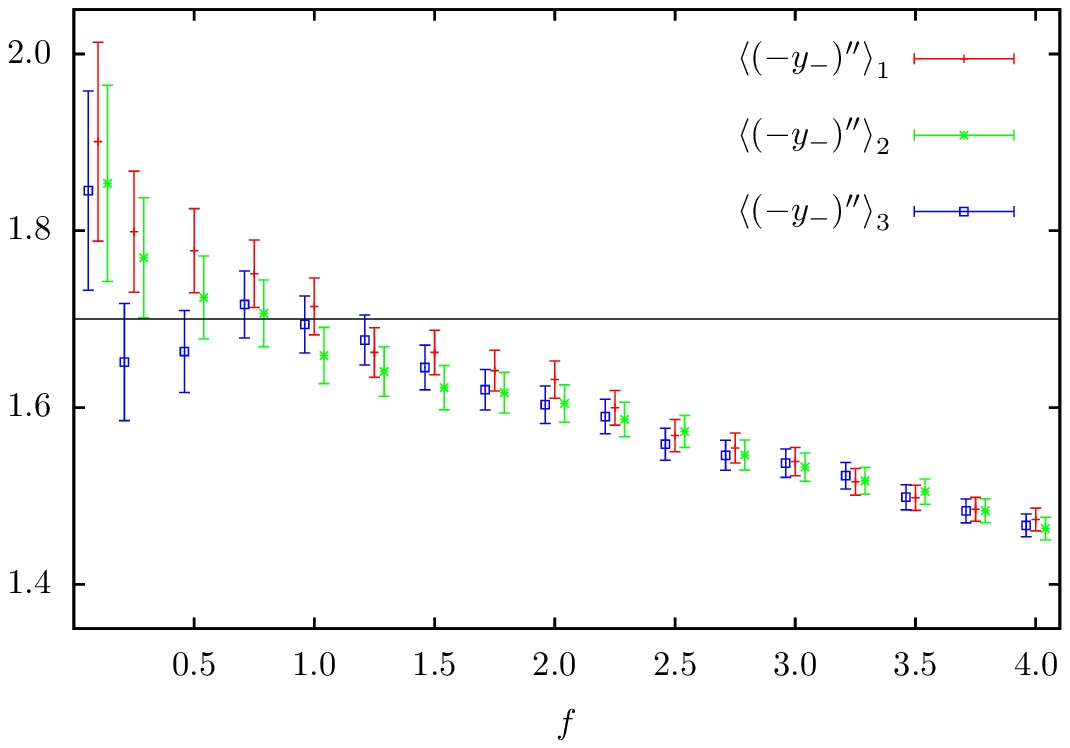}\hspace{12pt}\includegraphics[width=0.48\textwidth]{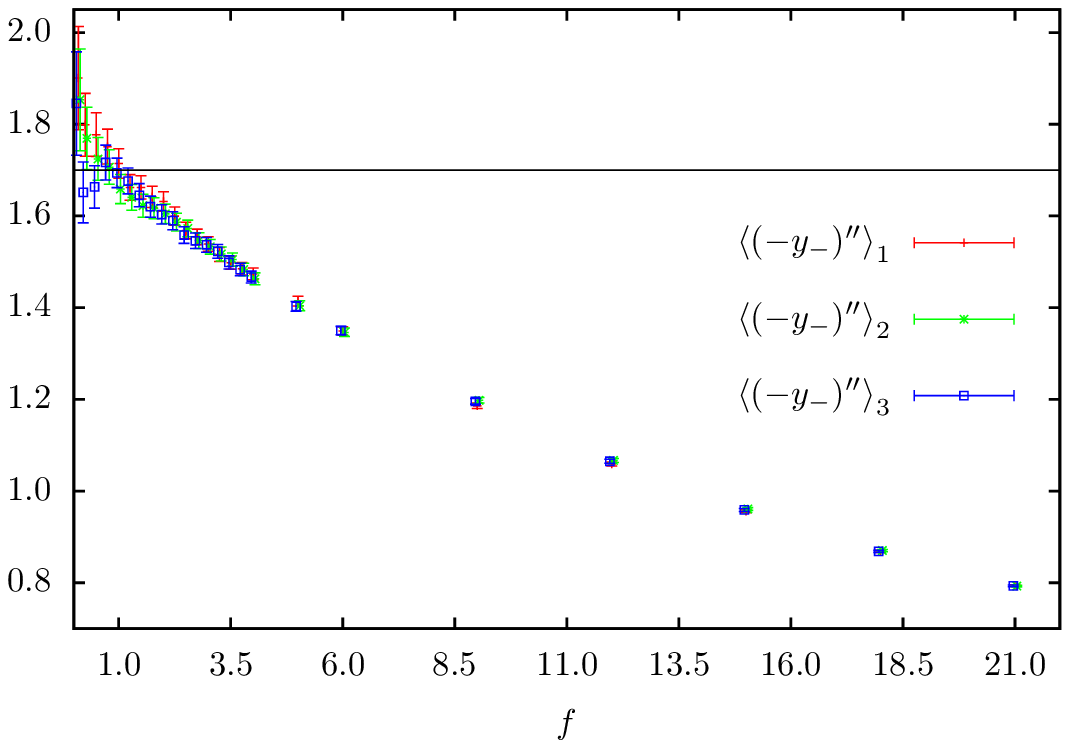}
\caption[Systematic errors of $\left\langle(-y_{-})^{\prime\prime}\right\rangle$ for $2^{10}$ points per loops]{Numerical values of $\left\langle(-y_{-})^{\prime\prime}\right\rangle$ as a function of the difference quotient parameter $f$ computed with three ensembles with $N=25\cdot10^{3}$ and $n_{ppl}=2^{10}$. The three sets of data points have been slightly shifted horizontally to be distinguishable in the plot. The solid black line is only for reference and comparison between different figures.\label{pic:Diff2-10-I}}
\end{figure}

\begin{figure}[!ht]
\includegraphics[width=0.48\textwidth]{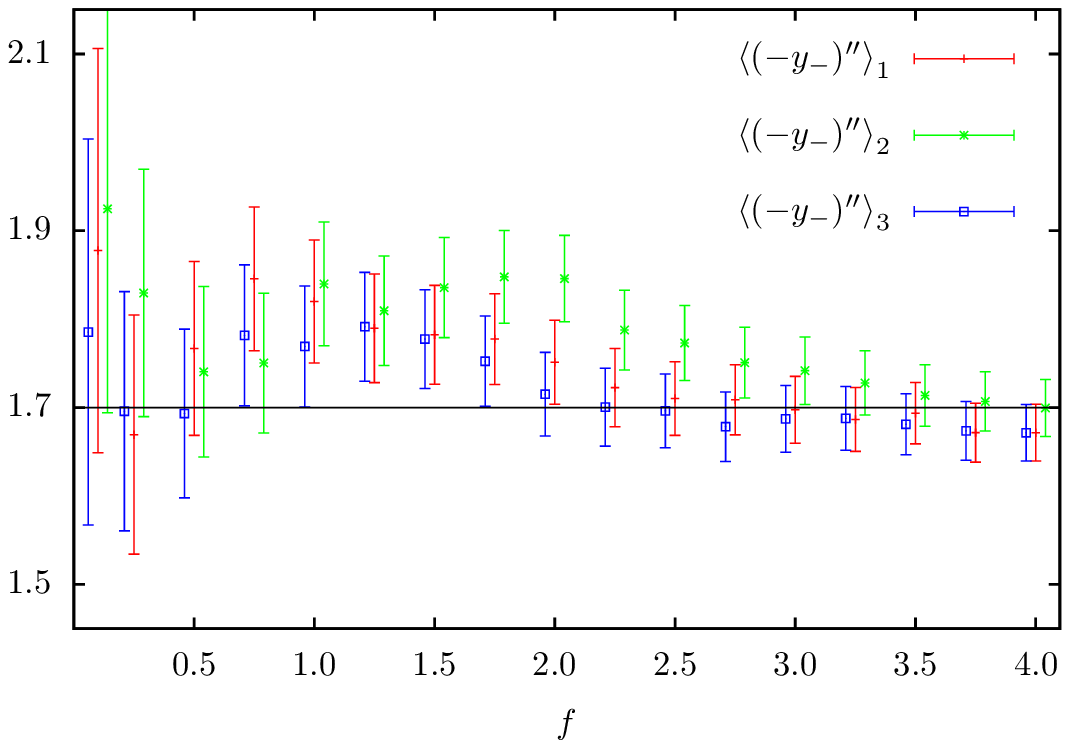}\hspace{12pt}\includegraphics[width=0.48\textwidth]{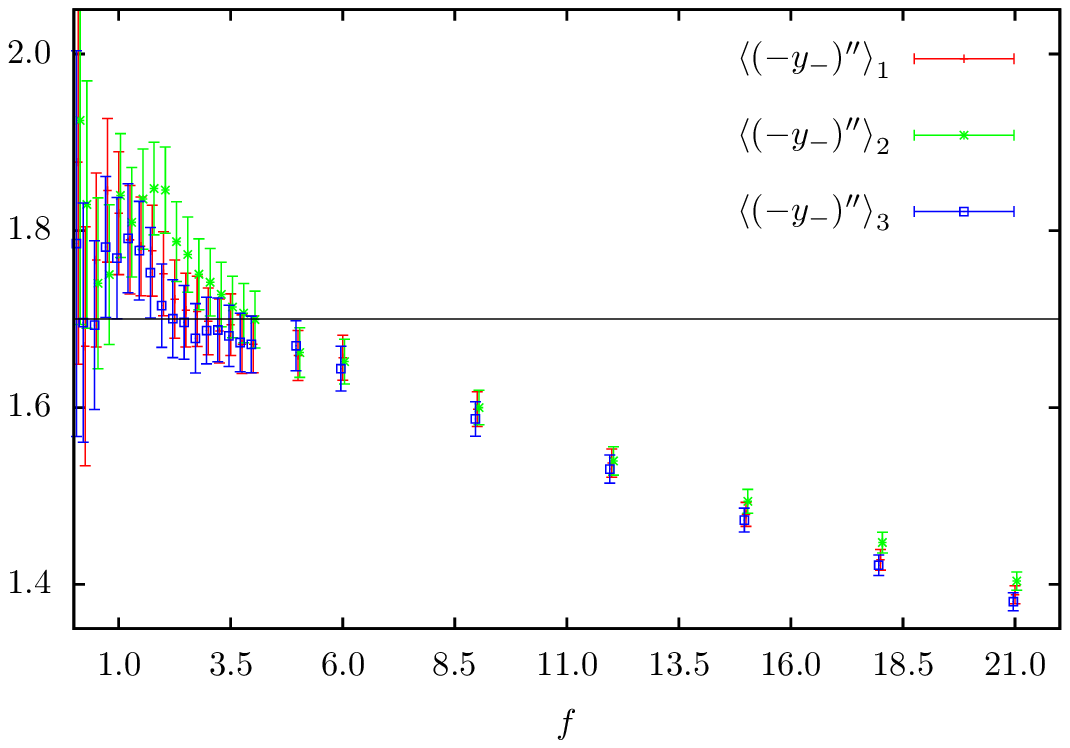}
\caption[Systematic errors of $\left\langle(-y_{-})^{\prime\prime}\right\rangle$ for $2^{14}$ points per loops]{Numerical values of $\left\langle(-y_{-})^{\prime\prime}\right\rangle$ as a function of the difference quotient parameter $f$ computed with three ensembles with $N=25\cdot10^{3}$ and $n_{ppl}=2^{14}$. The three sets of data points have been slightly shifted horizontally to be distinguishable in the plot. The solid black line is only for reference and comparison between different figures.\label{pic:Diff2-14-I}}
\end{figure}

\begin{figure}[!ht]
\includegraphics[width=0.48\textwidth]{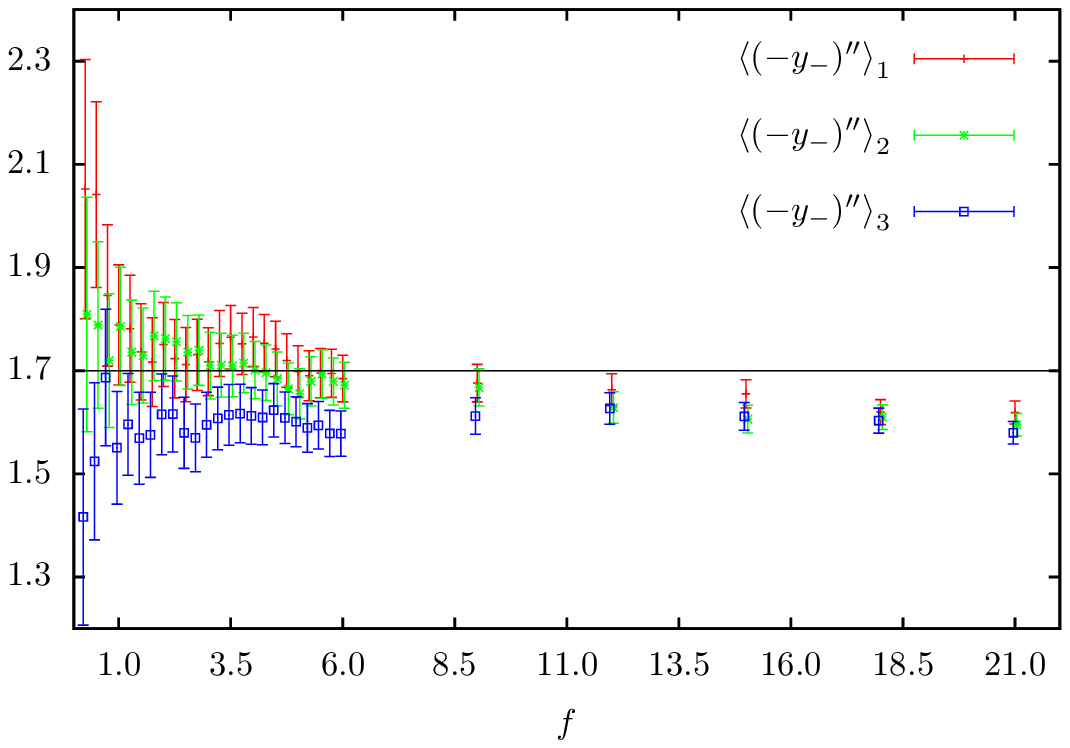}\hspace{12pt}\includegraphics[width=0.48\textwidth]{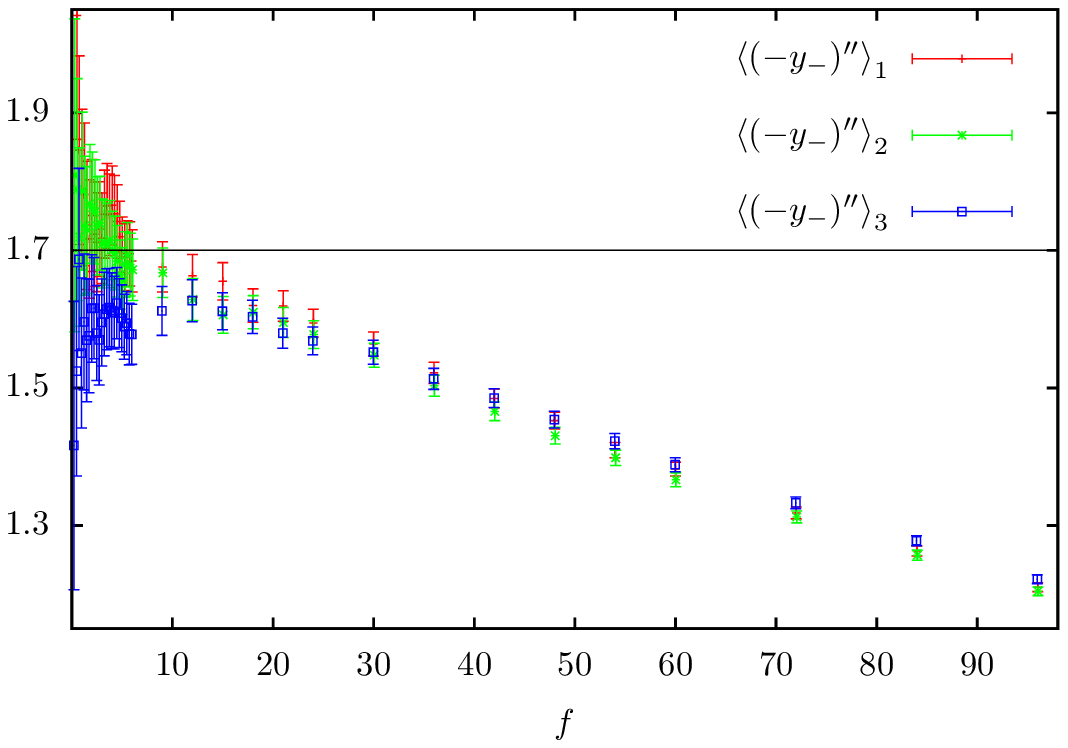}
\caption[Systematic errors of $\left\langle(-y_{-})^{\prime\prime}\right\rangle$ for $2^{17}$ points per loops]{Numerical values of $\left\langle(-y_{-})^{\prime\prime}\right\rangle$ as a function of the difference quotient parameter $f$ computed with three ensembles with $N=25\cdot10^{3}$ and $n_{ppl}=2^{17}$. The three sets of data points have been slightly shifted horizontally to be distinguishable in the plot. The solid black line is only for reference and comparison between different figures.\label{pic:Diff2-17-I}}
\end{figure}

\begin{figure}[!ht]
\includegraphics[width=0.48\textwidth]{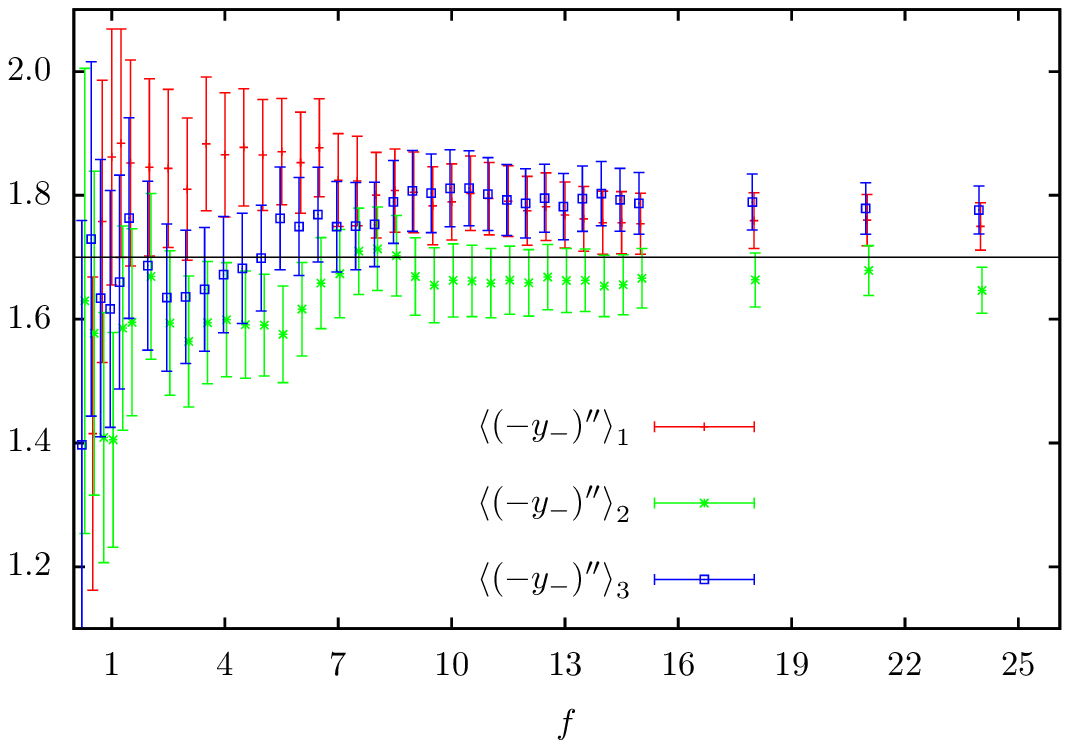}\hspace{12pt}\includegraphics[width=0.48\textwidth]{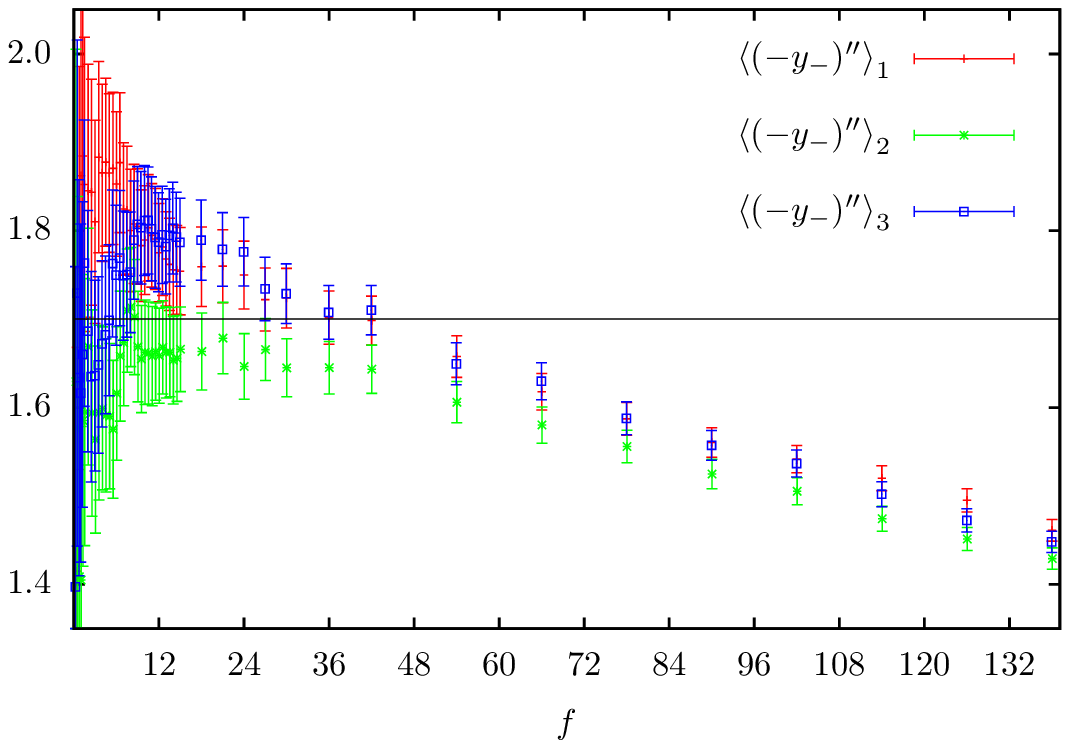}
\caption[Systematic errors of $\left\langle(-y_{-})^{\prime\prime}\right\rangle$ for $2^{20}$ points per loops]{Numerical values of $\left\langle(-y_{-})^{\prime\prime}\right\rangle$ as a function of the difference quotient parameter $f$ computed with three ensembles with $N=25\cdot10^{3}$ and $n_{ppl}=2^{20}$. The three sets of data points have been slightly shifted horizontally to be distinguishable in the plot. The solid black line is only for reference and comparison between different figures.\label{pic:Diff2-20-I}}
\end{figure}

The systematic errors in our calculations arise from two different
sources. The first arises from the discretization of the worldlines
themselves and is controlled by the number of points per lines
$n_{ppl}$. The continuum limit is defined by $n_{ppl}\to\infty$.
In \figref{pic:MP1Error} we show this by plotting our worldline
results for $\left\langle(-y_{-})^p\right\rangle$ with
$p\in\left\{1,2,3,4\right\}$ against $1/n_{ppl}$. For increasing
$n_{ppl}$, the numerical data approach the analytical values from
below. We fitted our
worldline data with a function $a+b/n_{ppl}$, disregarding the error
bars, i.e. the statistical errors, as they do not vary much for
different $n_{ppl}$. The systematic error is then the difference
between the analytical value and the fit in the limit
$n_{ppl}\to\infty$.
The second systematic error is introduced by the difference quotient used to compute the derivatives of the minimal point $\left\langle(-y_{-})\right\rangle$. For the numerator in \EQref{eq:YminAccel} to be non-vanishing, $h_\delta$ should be at least of the order of the average distance between points. Only then, the change of $t_{-}$ as a function of $\vec{\delta}$ can become visible; for too small $h_\delta$, $t_{-}$ simply does not vary such that $y^{\prime\prime}\to0$ arises as an artifact of the discretization.
At the same time, $h_\delta$ must be small compared to the extension of the entire loop. The variance of the points of the loop $\left\langle y^2\right\rangle\big\vert_{\delta=0}=1/6$ \cite{Gies:2003cv} gives a rough estimate of the extension of the loop. For the average distance between points, we use \begin{math} \left\langle\sum_{i=0}^{n_{ppl}}\left(y_{i+1}-y_{i}\right)^2/n_{ppl}\right\rangle^{1/2}\propto\sqrt{2/n_{ppl}}\end{math}. Both values are only rough measures, they become satisfactory approximations only for large $N$ and $n_{ppl}$. Combining both requirements, we find
\begin{align}
h_\delta = {} &\sqrt{\frac{2}{n_{ppl}}}\cdot f\ll\frac{1}{6}\quad\text{with } f\gtrsim\mathcal{O}(1)\ \ 
\Longrightarrow\ \  f = {} \frac{\varepsilon}{6}\cdot\sqrt{\frac{n_{ppl}}{2}}\quad\text{with } \varepsilon\ll1.\label{eq:fDef}
\end{align}
Since \EQref{eq:fDef} is just an estimate, we determine optimal values for $f$ directly by computing $\left\langle(-y_{-})^{\prime\prime}\right\rangle$ for different worldline ensembles with $N=25\cdot10^{3}$, $n_{ppl}=2^{10}\ldots2^{20}$ and several values $h_\delta=f \sqrt{2/n_{ppl}}$. We plot our results against $f$ (cf. \figref{pic:Diff2-10-I}-\ref{pic:Diff2-20-I}) and find large errors for small $f$, especially for $f\leq 1$. These values violate our requirement \EQref{eq:fDef}. For intermediate $f$ there is a plateau region beyond which the values of $\left\langle(-y_{-})^{\prime\prime}\right\rangle$ deviate (non-linearly). For these $f$ values a linearization is not a sufficient approximation for the derivative anymore.

\begin{figure}[!ht]
\centering
\includegraphics[width=0.48\textwidth]{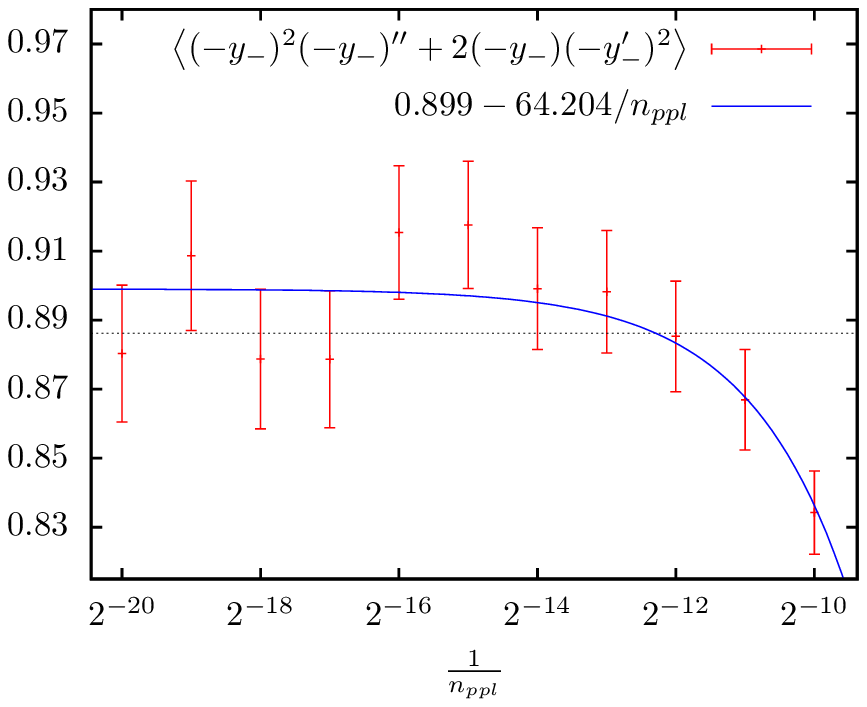}\hspace{12pt}\includegraphics[width=0.48\textwidth]{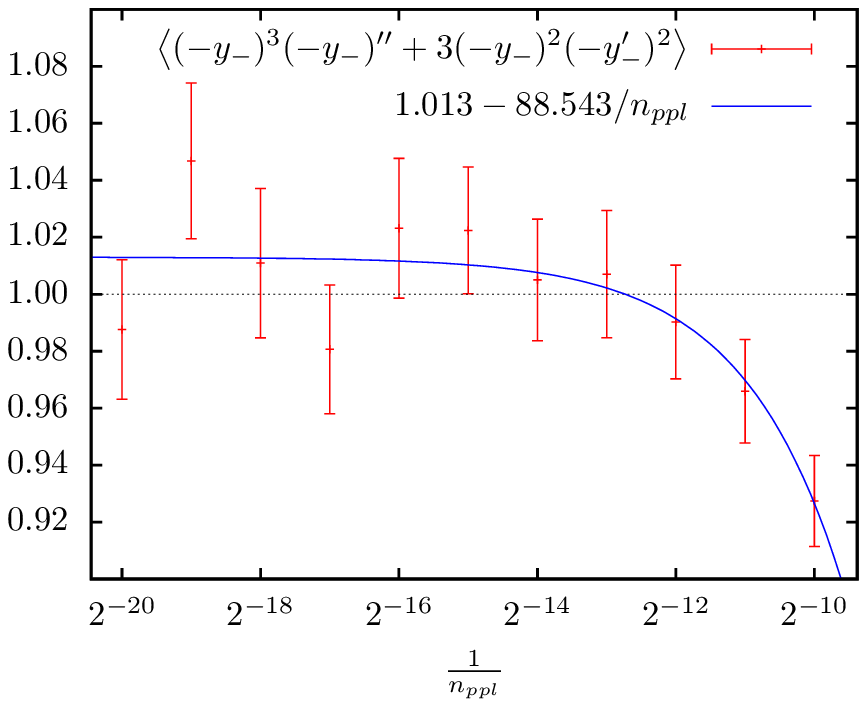}
\caption[Complete systematic errors for derivative terms]{Numerical values of the derivative terms in $d=2$ and $d=3$ for $n_{ppl}=2^{10}\ldots 2^{20}$, $N=25\cdot10^{3}$ and the corresponding $f$ from \tabref{TabValuesF}. The solid curves are fits to our data and the dotted lines are the respective analytical values. Statistical errors are given by the error bars, and the systematic errors correspond to the difference between analytical and fitted data in the limit $n_{ppl}\to\infty$.\label{pic:SysDiffErrors}}
\end{figure}

We read off optimal values for $f$ for different $n_{ppl}$ from the plateau region in \figref{pic:Diff2-10-I}-\ref{pic:Diff2-20-I}. These values are compiled in \tabref{TabValuesF}. We estimate the optimal $f$ for values of $n_{ppl}$ for which we have not run a calculation, such that the corresponding $\varepsilon$ are approximately $0.1$. Those values are consistent with the constraint in \EQref{eq:fDef}.
\begin{table}[!ht]
\begin{ruledtabular}
\begin{tabular}{ccccccc}
$n_{ppl}$ & allowed $f$ && allowed $\varepsilon$ && optimal $f$ & optimal $\varepsilon$ \\
$2^{10}$  & $1.0\ldots1.5$ && $0.265\ldots0.398$ && $1.00$ & $0.265$ \\
$2^{11}$  &  && && $1.00$ & $0.188$ \\
$2^{12}$  &  && && $1.25$ & $0.166$\\
$2^{13}$  &  && && $1.50$ & $0.141$\\
$2^{14}$  & $2.0\ldots4.0$ && $0.133\ldots0.265$ && $2.00$ & $0.133$ \\
$2^{15}$  &  && && $2.75$ & $0.130$\\
$2^{16}$  &  && && $3.50$ & $0.116$\\
$2^{17}$  & $2.5\ldots8.0$ && $0.059\ldots0.188$ && $4.00$ & $0.094$ \\
$2^{18}$  &  && && $6.00$ & $0.099$\\
$2^{19}$  &  && && $8.00$ & $0.094$\\
$2^{20}$  & $9.0\ldots25.0$ && $0.075\ldots0.207$ && $12.00$ & $0.099$\\
\end{tabular}
\end{ruledtabular}
\caption[Optimal values for $f$ and $\varepsilon$]{The optimal values for the algorithmic parameters $f$ and $\varepsilon$, respectively, in the range $n_{ppl}=2^{10}\ldots 2^{20}$. They are determined from \figref{pic:Diff2-10-I}-\ref{pic:Diff2-20-I} or estimated such that $\varepsilon\approx0.1$.\label{TabValuesF}}
\end{table}
In order to satisfy $f\gtrsim\mathcal{O}(1)$ and $\varepsilon\ll0$, our results suggest that computations of $\left\langle(-y_{-})^{\prime\prime}\right\rangle$ be only run with $n_{ppl}\geq2^{14}$ (cf. \figref{pic:Diff2-10-I} and \ref{pic:Diff2-14-I}). We deduce the systematic errors of averages of different powers of $(-y_{-})$ from the fits in \figref{pic:MP1Error}. Similarly we determine the systematic errors of the averages that contain derivatives of $(-y_{-})$ by computing these terms for different values of $n_{ppl}$ and the corresponding optimal value of $f$. \figref{pic:SysDiffErrors} shows the results of these calculations. For small $n_{ppl}$ our numerical data approach the analytical values from below, but are greater than the analytical values for large $n_{ppl}$. We again read off the systematic error as the difference between analytic values and our fit for $n_{ppl}\to\infty$.
This concludes our error discussion for the single Dirichlet plate. With the plots and optimal $f$ values from above, we are now able to run our algorithms with minimized errors.

To compute $\TnnCP$ and $\TzzCP$ with minimal errors we use an ensemble of $N=5\cdot10^{5}$ worldlines with $n_{ppl}=2^{20}$ points per line and choose $f=12$ ($\varepsilon=0.099$). The statistical errors are measured by the standard deviation, the systematic errors by the difference between analytical values and fits for $n_{ppl}\to\infty$. In \tabref{Tab1Plcomp} the worldline results are summarized for these optimized parameters.
\begin{table}[!ht]
\begin{ruledtabular}
\begin{tabular}{cccccc}
	                              & \text{analytic value}                  & \text{numerical value} & $\Delta_{\text{stat}}[\%]$ & $\Delta_{\text{sys}}[\%]$ & $\Delta_{\text{comp}}[\%]$ \\
$\left\langle(-y_{-})\right\rangle$   & $\frac{1}{2}\sqrt{\pi}\approx 0.88623$ & $0.88502 \pm0.00065$ & $0.073$ & $0.339$ & $0.412$\\
$\left\langle(-y_{-})^2\right\rangle$ & 1.00000                             & $0.99715 \pm 0.00141$ & $0.141$ & $0.500$ & $0.641$\\
$\left\langle(-y_{-})^3\right\rangle$ & $\frac{3}{4}\sqrt{\pi}\approx 1.32934$ & $1.32337 \pm 0.00289$ & $0.218$ & $0.527$ & $0.745$\\
$\left\langle(-y_{-})^4\right\rangle$ & $2.00000$                              & $1.98702 \pm 0.00626$ & $0.315$ & $0.550$ & $0.865$ \\
$\frac{1}{3}\left\langle\partial^2_{\delta_z}(-y_{-})^3\right\rangle$ & $\frac{1}{2}\sqrt{\pi}\approx0.88623$ & $0.89190 \pm 0.00452$ & $0.507$ & $1.467$ & $1.974$ \\
$\frac{1}{4}\left\langle\partial^2_{\delta_z}(-y_{-})^4\right\rangle$ & $1.00000$                             & $1.00103 \pm 0.00561$ & $0.560$ & $1.300$ & $1.860$ \\
\end{tabular}
\end{ruledtabular}
\caption[Optimized numerical results for the single plate]{Numerical results (and their standard deviations) for the single plate in $2$ and $3$ spatial dimensions with an ensemble of $5\cdot10^{5}$ worldlines, $2^{20}$ points per line and $h_\delta=12\,\sqrt{2/n_{ppl}}$. The statistical and systematic errors are shown explicitly, as well as the complete error $\Delta_{comp}$.\label{Tab1Plcomp}}
\end{table}
The statistical errors are well below $1\,\%$, due to the large number $N$ of worldlines in our ensemble, and the systematic errors are smaller than $1.5\,\%$. They are larger for higher powers of $(-y_{-})$. The errors of the derivative terms are generally larger, for they not only contain the systematic error for the point $(-y_{-})$, determined by $n_{ppl}$, but also the error due to the linearization of the derivatives, determined by $h_\delta$. To summarize, the numerical worldline method is capable of computing the relevant EMT components for the present case on a satisfactory percent precision level.

\section{The energy-momentum tensor for parallel plates\label{sec:CasPl}}

Our second proof-of-principle example is the case of two parallel
plates with Dirichlet boundary conditions. This configuration
corresponds to the original setup studied by Casimir for the
electromagnetic field \cite{Casimir}. The EMT in this setting has 
been studied in the literature, see, e.g.,
\cite{Milton:2002vm}. As in the previous section, we compute $\Tnn$ and
$\Tzz$ in $d=2$ and $d=3$ dimensions. With these terms, we can see if
the null energy condition is violated or fulfilled. The plates are
placed at $z=a/2$ and $z=-a/2$ and the $z$ axis is perpendicular to
them. The whole configuration is invariant under arbitrary
translations parallel to the plates.

\subsection{Analytical calculation for two parallel plates}

$T_{00}(\vec{x},t)$ for two parallel Dirichlet plates can be calculated analytically by applying the method of images and using the results of \secref{sec:AnaSinglePlate} for the Green's functions. The two plates at $z=\pm a/2$ constitute the boundary and decompose the domain $\mathcal{D}$ into three disjoint regions. In the outside regions $z>a/2$ and $z<-a/2$, we find the Green's function for a single plate at $z=a/2$ and $z=-a/2$, respectively. The Green's function between the plates is given by the method of images as an infinite series of image charges induced on the plates by a point source. For arbitrary $\vec{x}$ and $\vec{x}^{\,\prime}$ between the plates for which $\vert z\vert<a/2$ and $\vert z^{\,\prime}\vert<a/2$, we introduce $\vec{x}_q:=\left(\vec{x}_{||},z+2qa\right)$ and $\widetilde{\vec{x}}_q:=\left(\vec{x}_{||},-z+\left(2q+1\right)a\right)$. The Green's functions $\Geff=\Gsigma-\Gzero$ are then
\begin{align}
\begin{split}
\Geff\big\vert^{d=2} = {} &
\frac{1}{2\pi}\sum_{q\in\,\mathbb{Z}}^\infty\left(\mathcal{K}_0(-ik\vert\vec{x}_q-\vec{x}^{\,\prime}\vert)-\mathcal{K}_0(-ik\vert\widetilde{\vec{x}}_q-\vec{x}^{\,\prime}\vert)\right)
-\frac{1}{2\pi}\mathcal{K}_0(-ik\vert\vec{x}_{q=0}-\vec{x}^{\,\prime}\vert),\\
\Geff\big\vert^{d=3} = {} &
\frac{1}{4\pi}\sum_{q\in\,\mathbb{Z}}^\infty\left(\frac{\er^{ik\vert\vec{x}_q-\vec{x}^{\,\prime}\vert}}{\vert\vec{x}_q-\vec{x}^{\,\prime}\vert}-\frac{\er^{ik\vert\widetilde{\vec{x}}_q-\vec{x}^{\,\prime}\vert}}{\vert\widetilde{\vec{x}}_q-\vec{x}^{\,\prime}\vert}\right)
-\frac{1}{4\pi}\frac{\er^{ik\vert\vec{x}_{q=0}-\vec{x}^{\,\prime}\vert}}{\vert\vec{x}_{q=0}-\vec{x}^{\,\prime}\vert}.\nonumber
\end{split}
\end{align}
After inserting these Green's functions into \EQref{eq:EMTG}, we integrate with respect to $k$. We simplify the result further by assuming that $\vec{x}$ and $\vec{x}^{\,\prime}$ lie on the $z$ axis, that is, $\vec{x}_{||}=\vec{x}_{||}^{\,\prime}$. The various EMT components can now be written in the form of a series \begin{math}\zeta(s,f(z_{cp},\Delta_z)) {} := \sum_{q=0}^\infty\left(f(z_{cp},\Delta_z)+q\right)^{-s}\end{math}, which is a Hurwitz $\zeta$ function for $\ReP\left(s\right)>1$ and $\ReP\left(f(z_{cp},\Delta_z)\right)>0$. While we always have $f(z_{cp},\Delta_z)>0 $, for $d=2$ we find $s=1$ in $\TzzCP$. We solve this problem by first acting with the derivatives $\partial_{z_{cp}}$ and $\partial_{\Delta_z}$, which increases the exponent to $s=3$.

In $d=2$ the EMT between the plates, $\vert z_{cp}\vert<a/2$, is given by the components 
\begin{subequations}
\begin{align}
\TnnCP\big\vert_{I}^{d=2} {}\hspace{0.6em} = & -\frac{1}{32\pi a^{\,3}}\left[2\zeta(3)-\zeta\left(3,\frac{1}{2}+\frac{z_{cp}}{a}\right)-\zeta\left(3,\frac{1}{2}-\frac{z_{cp}}{a}\right)\right], \\
\TnnCP\big\vert_{II}^{d=2} {}\hspace{0.6em} = & -\frac{1}{16\pi a^{\,3}}\left[\zeta\left(3,\frac{1}{2}+\frac{z_{cp}}{a}\right)+\zeta\left(3,\frac{1}{2}-\frac{z_{cp}}{a}\right)\right] ,\\
\TzzCP\big\vert_{Ia}^{d=2} {}\hspace{0.6em} = & -\frac{1}{16\pi a^{\,3}}\left[\zeta\left(3,\frac{1}{2}+\frac{z_{cp}}{a}\right)+\zeta\left(3,\frac{1}{2}-\frac{z_{cp}}{a}\right)\right] ,\\
\TzzCP\big\vert_{Ib+Ic}^{d=2} {} = & -\frac{1}{16\pi a^{\,3}}\,2\zeta(3).
\end{align}
\end{subequations}
The corresponding 3-dimensional results are
\begin{subequations}
\begin{align}
\TnnCP\big\vert_{I}^{d=3} {}\hspace{0.6em} = & -\frac{1}{32\pi^2 a^{\,4}}\left[2\zeta(4)-\zeta\left(4,\frac{1}{2}+\frac{z_{cp}}{a}\right)-\zeta\left(4,\frac{1}{2}-\frac{z_{cp}}{a}\right)\right], \\
\TnnCP\big\vert_{II}^{d=3} {}\hspace{0.6em} = & -\frac{3}{32\pi^2 a^{\,4}}\left[\zeta\left(4,\frac{1}{2}+\frac{z_{cp}}{a}\right)+\zeta\left(4,\frac{1}{2}-\frac{z_{cp}}{a}\right)\right] ,\\
\TzzCP\big\vert_{Ia}^{d=3} {}\hspace{0.6em} = & -\frac{3}{32\pi^2 a^{\,4}}\left[\zeta\left(4,\frac{1}{2}+\frac{z_{cp}}{a}\right)+\zeta\left(4,\frac{1}{2}-\frac{z_{cp}}{a}\right)\right] ,\\
\TzzCP\big\vert_{Ib+Ic}^{d=3} {} = & -\frac{3}{32\pi^2 a^{\,4}}\,2\zeta(4).
\end{align}
\end{subequations}
As before $\TzzCP\big\vert_{Ib}$ can be computed using $\TnnCP\big\vert_{I}=-\TzzCP\big\vert_{Ic}$. The EMT components diverge as the distance from either plate goes to zero.
The NEC along the $z$ axis between the plates is also violated in the case of two parallel plates,
\begin{align}
T_{00} + T_{zz} {} = & \left\{\renewcommand{\arraystretch}{1.0}\begin{array}{cc} \mbox{{$-\left[6\zeta(3)+\zeta\left(3,\frac{1}{2}+\frac{z_{cp}}{a}\right)+\zeta\left(3,\frac{1}{2}-\frac{z_{cp}}{a}\right)\right]/\left(32\pi a^{\,3}\right)$}} & \mbox{\ for $d=2$}\\ \mbox{{$-\left[4\zeta(4)+\zeta\left(4,\frac{1}{2}+\frac{z_{cp}}{a}\right)+\zeta\left(4,\frac{1}{2}-\frac{z_{cp}}{a}\right)\right]/\left(16\pi^2 a^{\,4}\right)$}} & \mbox{\ for $d=3$}\end{array}\renewcommand{\arraystretch}{1}\right..
\end{align}
Our results agree with those of \cite{Milton:2002vm}.

\subsection{The conformal complement of $\EMTOp$}

The energy-momentum tensor for the electromagnetic field with perfect conductor boundary conditions does not between the plates \cite{Brown:1969na}. However, the electromagnetic EMT has zero trace because the electromagnetic field is conformally invariant. The energy-momentum tensor derived for a conformally coupled scalar field between two parallel plates with Dirichlet boundary conditions does also not diverge near the plates but it still yields the same total energy and momentum as for the minimally coupled scalar. The difference between both tensors is the conformal complement \begin{math} \Delta\EMTOp {} := -\xi\left(\partial_{\mu}\partial_{\nu}-g_{\mu\nu}\partial_{\alpha}\partial^{\alpha}\right)\fieldPhi\fieldPhi\end{math}, where $\xi=(d-1)/(4d)$ \cite{CallanJr197042}. Adding the vacuum expectation value of the components of this conformal complement operator to our canonical EMT removes the divergent terms of the latter. We determine 
the conformal complement in the worldline formalism along the lines of the calculations performed for $\Tnn$ and $\Tzz$ in \secref{sec:comp_op}. We find 
\begin{math}\Delta\Tnn {} = -4\xi\cdot\Tnn\big\vert_{II}\end{math} and \begin{math}\Delta\Tzz {} = -4\xi\Tzz\big\vert_{Ia}+4\xi\Tnn\big\vert_{II}\end{math}. As a consequence, the conformal or improved energy-momentum tensor $\Theta_{\mu\nu}(\vec{x},t):=T_{\mu\nu}(\vec{x},t)+\Delta T_{\mu\nu}(\vec{x},t)$ is then finite, even constant, but still violates the NEC:
\begin{align}
\begin{split}
\Theta_{00}(\vec{x},t)+\Theta_{zz}(\vec{x},t) {} = & -\frac{1}{32\pi^{\ } a^{\,3}}\,2\zeta(3)-2\frac{1}{32\pi^{\ } a^{\,3}}\,2\zeta(3)\ \text{\ \ for $d=2$},\\
\Theta_{00}(\vec{x},t)+\Theta_{zz}(\vec{x},t) {} = & -\frac{1}{32\pi^2 a^{\,4}}\,2\zeta(4)-3\frac{1}{32\pi^2 a^{\,4}}\,2\zeta(4)\ \text{\ \ for $d=3$}.
\end{split}
\end{align}
Our results for $\Theta_{\mu\nu}(\vec{x},t)$ match with those of \cite{Milton:2002vm} where the general case of a massive scalar field with Dirichlet BCs in $d$ spatial dimensions is presented. Since we can always construct $\Theta_{\mu\nu}(\vec{x},t)$ from its canonical counterpart $T_{\mu\nu}(\vec{x},t)$, we restrict our computations to the latter.

\subsection{Worldline calculation for two parallel Dirichlet plates}

\begin{figure}[!ht]
\centerline{\includegraphics[width=0.85\textwidth]{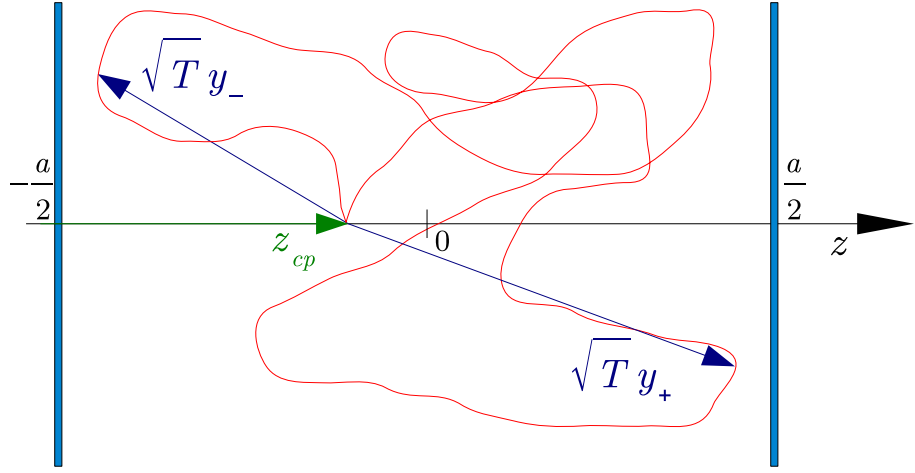}}
\caption[Setup of Casimir's parallel plates]{Sketch of the Casimir plates setup with an exemplary unit loop. \label{pic:CasimirPlates}}
\end{figure}

We start our worldline calculation with identifying the intersection condition from \figref{pic:CasimirPlates}.
The worldlines have now two possibilities to intersect the plates, which means we need to identify the two extremal points $y_{-}$ and $y_{+}$ of each worldline. More precisely, we need to determine which point intersects one of the plates first. For any given line this will depend on the position $z_{cp}$ on the $z$ axis. The intersection condition $\mathcal{M}$ in $\Theta\left(\sqrt{T}\mathcal{M}-1\right)$ is then
\begin{align}\label{eq:IntCondCasPlates}
\left.
\begin{array}{ccc}
\sqrt{T}y_{-}+z_{cp} & \leq & -\frac{a}{2}\\[0.5em]
\sqrt{T}y_{+}+z_{cp} & \geq & \ \frac{a}{2}  
\end{array}
\right\}
 & \Longrightarrow\ \text{with  }\ \mathcal{M} = {} \Max\left(\mathcal{M}_{+},\mathcal{M}_{-}\right) := {} \Max\left[\frac{(y_{+})}{\frac{a}{2}-z_{cp}},\frac{(-y_{-})}{\frac{a}{2}+z_{cp}}\right].
\end{align}
The condition in \EQref{eq:IntCondCasPlates} is 1-dimensional, so that 1-dimensional worldlines suffice for the computation. However, the calculation, especially of the derivatives of $\mathcal{M}$, is more involved because $\mathcal{M}$ is the maximum function \begin{math}\mathcal{M} = {} \left(\mathcal{M}_{+}+\mathcal{M}_{-}+\left\vert\mathcal{M}_{+}-\mathcal{M}_{-}\right\vert\right)/2\end{math}. Due to the modulus, $\mathcal{M}$ -- if taken at face value -- is not 
differentiable at $\mathcal{M_{+}}=\mathcal{M_{-}}$, neither in $z_{cp}$ nor in $\delta_z$. As before, 
we still treat $\mathcal{M}$ as if it were differentiable in order to identify a constructive definition of the derivative inside the path integral in terms of geometric properties of the worldline. 
Worldline averages of powers of $\mathcal{M}$ are straightforward to compute. Averages of first derivatives of $\mathcal{M}$ also pose no difficulties because $\partial_x\vert x\vert=\Sign(x)$. 
However, worldline averages of $\partial^2\mathcal{M}$ involve the derivative of the $\Sign$ function. Formally, one has $\partial_x\Sign\left(x\right)\propto\delta\left(x\right)$ whose numerical evaluation under the worldline average is not straightfoward.
So far, we have not been able to find a suitably smeared-out $\delta$ function adapted to the numerical discretization with controllable errors. 
Instead, we simply 
approximate $\partial_x\Sign\left(x\right)$ with a difference quotient. In the discretized numerical framework, there is no unique variable choice. 
For example, one can consider
$\Sign\left[\mathcal{M}_{+}-\mathcal{M}_{-}\right]$ as a function of
one variable $\mathcal{M}_{+}-\mathcal{M}_{-}$ or as a function of the
two variables $\mathcal{M}_{+}$ and $\mathcal{M}_{-}$. The difference
quotient of either option suffers from problems. In the first version
$\mathcal{M}_{+}-\mathcal{M}_{-}$ can go to zero and will
na{\"i}vely generate diverging terms. The second version, while
avoiding divergences and yielding better results, still has a large
systematic error. A more successful strategy arises from a third
option to view the $\Sign$ function as a function of
$\mathcal{M}_{+}/\mathcal{M}_{-}$. This is possible because both
$\mathcal{M_{+}}$ and $\mathcal{M_{-}}$ are positive. The resulting
difference quotient is thus written as
\begin{align}
\partial\Sign\left[\mathcal{M}_{+}-\mathcal{M}_{-}\right]
= {} & \partial\Sign\left[\frac{\mathcal{M}_{+}}{\mathcal{M}_{-}}-1\right] = {} \partial_{\frac{\mathcal{M}_{+}}{\mathcal{M}_{-}}}\Sign\left[\frac{\mathcal{M}_{+}}{\mathcal{M}_{-}}-1\right]\,\partial\,\frac{\mathcal{M}_{+}}{\mathcal{M}_{-}}\nonumber\\[0.5em]
= {} & \frac{\Sign\left[\frac{\mathcal{M}_{+}}{\mathcal{M}_{-}}\left(1+l\right)-1\right]-\Sign\left[\frac{\mathcal{M}_{+}}{\mathcal{M}_{-}}\left(1-l\right)-1\right]}{2l}\frac{\mathcal{M}_{-}}{\mathcal{M}_{+}}\Bigg\vert_{l\to0}\ \partial\,\frac{\mathcal{M}_{+}}{\mathcal{M}_{-}}.\label{eq:DiffSgn}
\end{align}
In \EQref{eq:DiffSgn} $l$ must satisfy $l\ll 1$ but must not be so small as to render the numerator of the difference quotient equal to zero for too many lines in the ensemble. The systematic error introduced by this linearization of a derivative is controlled by $l$. We note that if either $\mathcal{M}_{+}$ or $\mathcal{M}_{-}$ should be zero then the $\partial\Sign$ term vanishes. However, this can only happen if either $y_{-}$ or $y_{+}$ is equal to zero. Although such worldlines exist, their number decreases with increasing $n_{ppl}$, i.e., they form a set of measure zero in the worldline integral.

\subsection{Numerical results for $\TnnCP$ and $\TzzCP$}

Most sources of errors are already known. The statistical error is related to the number $N$ of lines in our ensemble and the number of points per line $n_{ppl}$ determines a systematic error of the Monte Carlo integration. Furthermore, $h_\delta$ controls the systematic error of the derivatives $(\pm y_{\pm})^\prime$ and $(\pm y_{\pm})^{\prime\prime}$. Since the worldline ensemble for closed loops is spherically symmetric in the continuum limit, the values of $\left\langle(-y_{-})\right\rangle$ and $\left\langle(y_{+})\right\rangle$ are identical and so are their error estimates. Therefore, we can immediately use the optimized values of these three parameters that we found for the single plate configuration.

\begin{figure}[!t]
\includegraphics[width=0.48\textwidth]{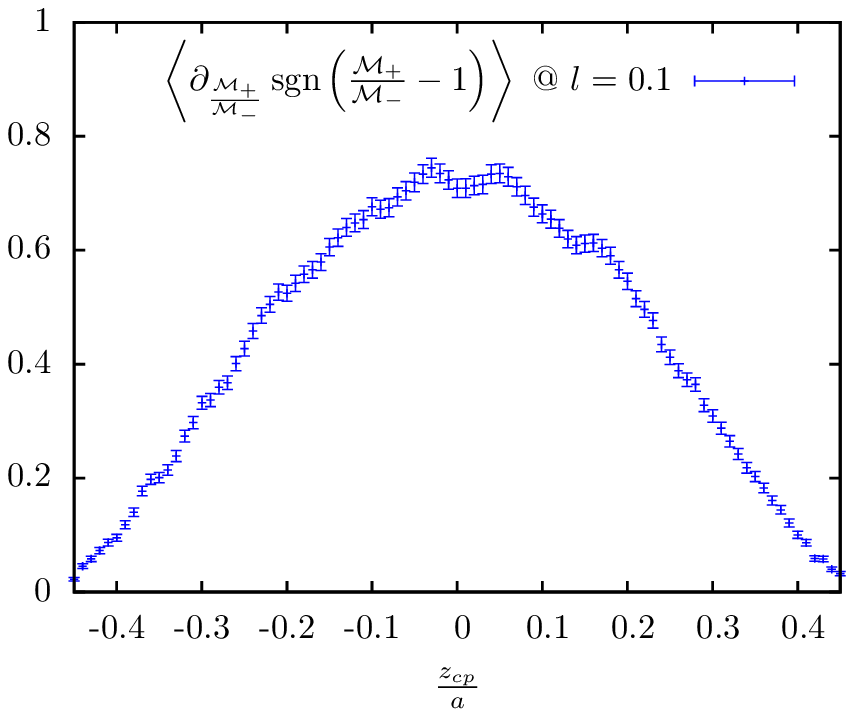}\hspace{12pt}\includegraphics[width=0.48\textwidth]{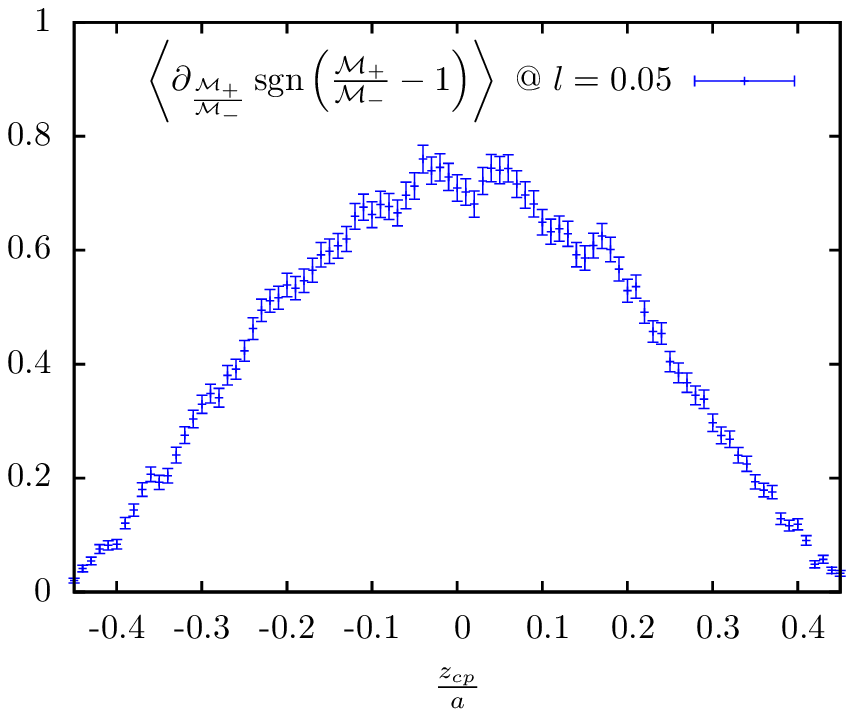}\\
\includegraphics[width=0.48\textwidth]{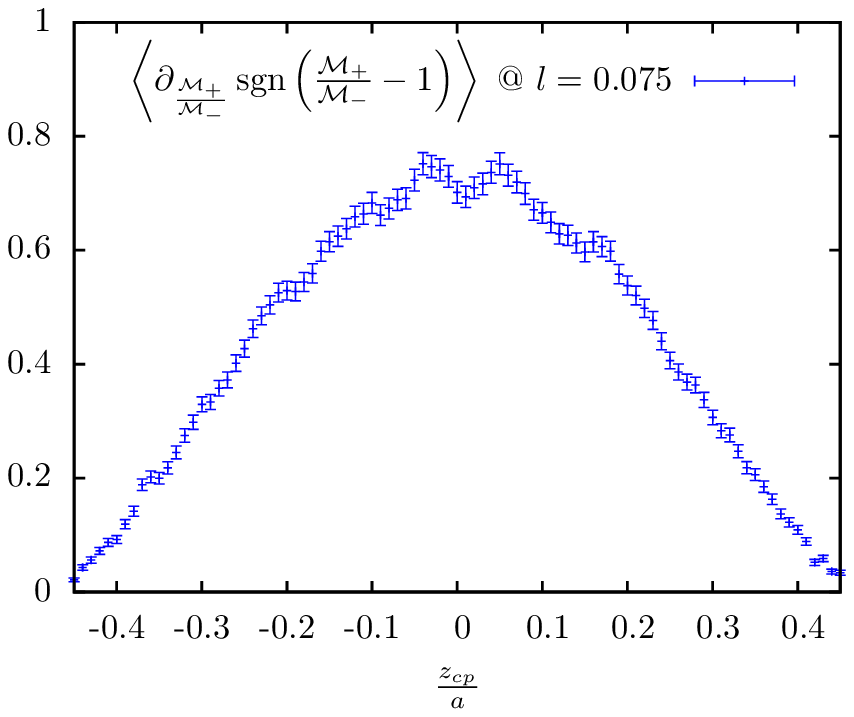}\hspace{12pt}\includegraphics[width=0.48\textwidth]{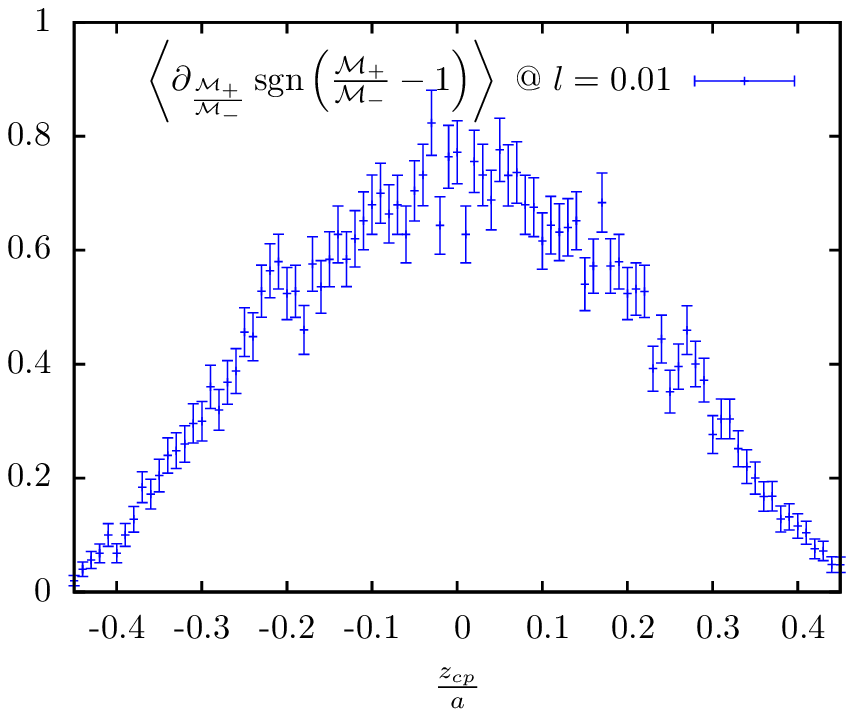}
\caption[Error estimate of the linearization of $\partial\Sign$]{Worldline average of the derivative of the sign function in between the plates for different values of the linearization parameter $l$. Fluctuations and errors are smallest for $l=0.075$.\label{pic:SigError}}
\end{figure}

The linearization of $\partial\Sign$ in $\partial^2\mathcal{M}$ introduces yet another systematic error that is determined by $l$. In order to find an optimal value of $l$, we compute the expectation value of $j(z_{cp}/a)=\partial_x\Sign(x-1)$ with $x=\mathcal{M_{+}}/\mathcal{M_{-}}$ as a function of $z_{cp}/a$. For that, we evaluate the difference quotient in \EQref{eq:DiffSgn} for different values of $l\ll1$. We show the results in \figref{pic:SigError}. The function $j(z_{cp}/a)$ is positive with large values around $z_{cp}=0$ and decreases to zero near the plates. $j(z_{cp}/a)$ is only non-zero, if a worldline switches from intersecting one plate to intersecting the other plate first. This happens more often when the worldlines are in the middle than when they are close to one plate. For $l=0.1$ and $l=0.075$ the average of the derivative becomes a smooth function with small statistical errors. For smaller $l$ the errors increase and the fluctuations around the average value become large. This is due to the fact that for too small $l$ the numerator of the difference quotient tends to zero for too many worldlines in the average. We choose $l=0.075$ as an optimal value. The remaining errors are reduced by an increase of $n_{ppl}$ and $N$.

\begin{figure}[!ht]
\includegraphics[width=0.48\textwidth]{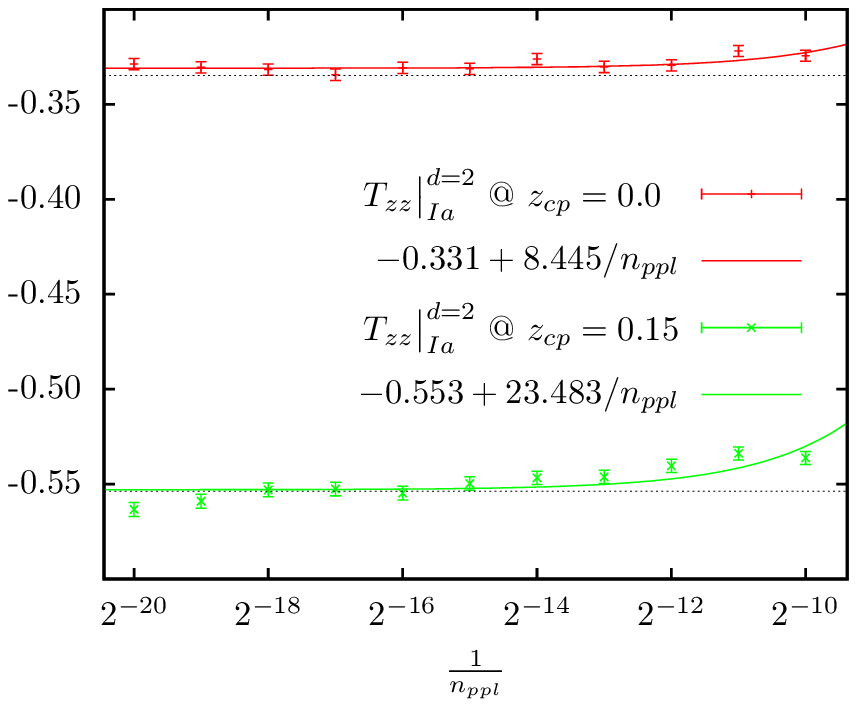}\hspace{12pt}\includegraphics[width=0.48\textwidth]{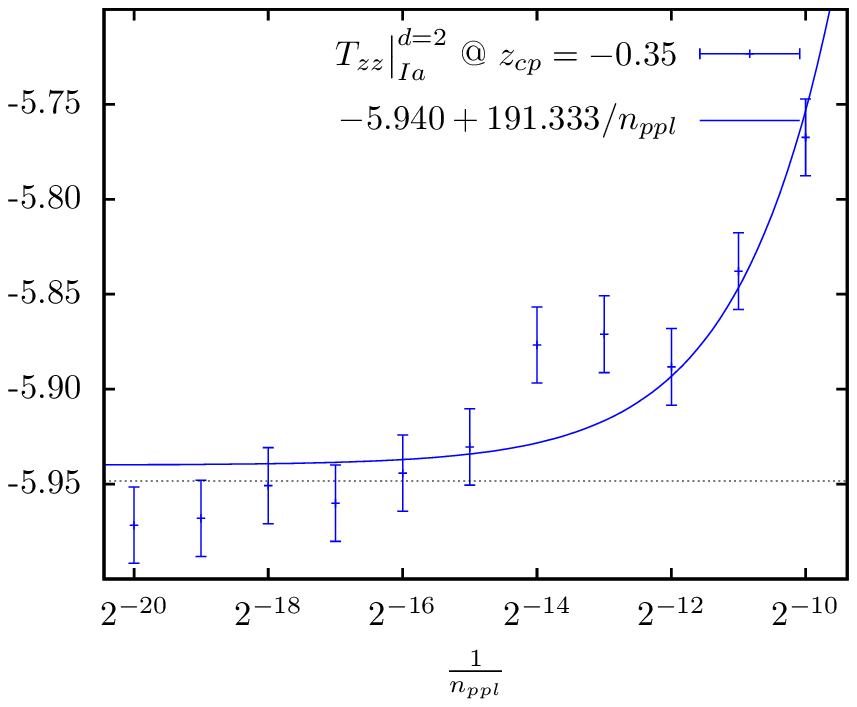}
\caption[Systematic errors of EMT components in the parallel plates
  configuration in $d=2$ dimensions]{Numerical values of
  $T_{zz}\big\vert_{Ia}$ in $d=2$ dimensions at $z_{cp}\in\{0.15,\,
  0.0\, ,-0.35\}$ as a function of $1/n_{ppl}$. The statistical errors
  are represented by the error bars, and systematic errors are given
  by the difference between the fits to our data (solid lines) and the
  analytical values (dotted lines) in the limit
  $n_{ppl}\to\infty$. The plots for the other EMT components, as well
  as those for $d=3$, are qualitatively the
  same.\label{pic:CasTmunu2dSysError}}
\end{figure}

\begin{figure}[!ht]
\includegraphics[width=0.80\textwidth]{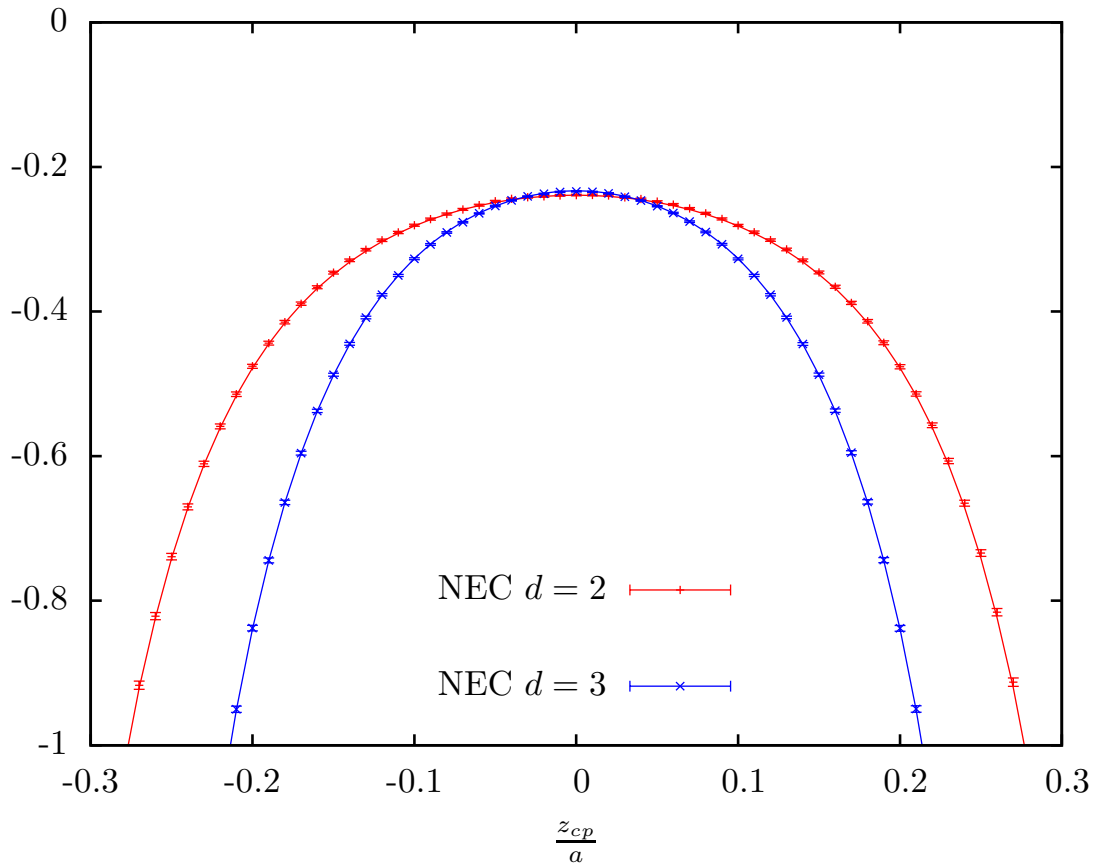}
\caption[NEC along the $z$ axis for two parallel plates in 2 and 3 spatial dimensions]{Results for 
the null energy condition (NEC) for two parallel Dirichlet plates in 2 and 3 spatial dimensions. It is violated at every point in between the plates and diverges near them. The worldline numerical result (points with error bars) matches very well with the analytical result (solid lines).\label{pic:CasimirNEC}}
\end{figure}

We ascertain the overall systematic errors by determining how $\TnnCP$ and $\TzzCP$ converge to their analytic values for increasing $n_{ppl}$. In \figref{pic:CasTmunu2dSysError} we show numerical values for one exemplary component of the energy-momentum tensor for three different $z_{cp}$, using ensembles with $N=25\cdot 10^{3}$, $n_{ppl}\in\{10,\ldots,20\}$ and the corresponding optimal value of $h_\delta$ and $l=0.075$. We again fit our data to a function $a+b/n_{ppl}$ such that $a$ estimates the remaining systematic error for $n_{ppl}\to\infty$. We display those error estimates in \tabref{tab:CasSysError}.
The systematic errors vary with $z_{cp}$ and so we take the largest error, $2.5\,\%$, as a general estimate.
\begin{table}[!ht]
\begin{ruledtabular}
\begin{tabular}{lccccccccc}
&&&& $d=2$&&&& $d=3$ &\\
&$z_{cp}$ & & $0.15$ & $0.0$ & $-0.35$ & & $0.15$ & $0.0$ & $-0.35$ \\
$\Delta_{sys}[\%]$
&$\TnnCP\big\vert_{I}$ &  & $1.2$ & $0.7$ & $0.6$ & & $1.8$ & $1.0$ & $0.8$\\
$\Delta_{sys}[\%]$
&$\TzzCP\big\vert_{Ia}$ &  & $0.2$ & $1.2$ & $0.1$ & & $0.6$ & $0.6$ & $0.4$\\
$\Delta_{sys}[\%]$
&$\TzzCP\big\vert_{Ib}$ &  & $1.5$ & $1.0$ & $1.4$ & & $2.5$ & $0.5$ & $0.5$\\
\end{tabular}
\end{ruledtabular}
\caption[Systematic errors for the original Casimir setup]{Systematic errors $\Delta_{sys}$ for the original Casimir setup, two parallel plates, with ensembles of $N=25\cdot 10^{3}$ worldlines at $z_{cp}\in\{0.15,\, 0.0\, ,-0.35\}$. These values are derived as the difference between fits to our numerical data and the analytical values in the limit $n_{ppl}\to\infty$ (cf. \figref{pic:CasTmunu2dSysError}). We use the largest error of $2.5\,\%$ as a general estimate.\label{tab:CasSysError}}
\end{table}
To arrive at results with minimal errors we increase the number of loops to $N=5\cdot10^{5}$ and the number of points per loop to $n_{ppl}=2^{20}$. This value corresponds to $f=12$ and $\varepsilon=0.099$ according to \tabref{TabValuesF}. We also choose $l=0.075$. \figref{pic:CasimirNEC} shows our optimized numerical data in comparison with the analytical result for the null energy condition in $d=2$ and $d=3$ dimensions. In both cases the null energy condition is violated at every point in between the plates with divergences near the plates. The differences between analytical and numerical values are satisfactorily small with statistical errors smaller than $0.6\,\%$ and the general estimate of $2.5\,\%$ systematic error.

\section{Conclusions}

We have shown how the worldline formalism can be applied to local
operators like the energy-momentum tensor of a scalar field that is
coupled to a static background potential $\potential$. The general
properties for computations of the EMT are similar to those of quantum
actions or energies: the fluctuations are represented by spacetime
trajectories of the fluctuating particles parametrized by an intrinsic
propertime the total amount of which is a measure for the
spatiotemporal extent of the fluctuation. Information about
fluctuation-induced quantities is thus encoded in the geometric
properties of the worldlines relative to the Casimir configuration,
such as intersection properties of the worldlines with the plates. The
essential difference for EMT computations is that in addition to
closed trajectories also open trajectories have to be considered as a
consequence of the point-splitting definition of the composite
operator. While closed worldlines are sufficient for the energy
component $T_{00}$, the stress components also receive contributions
from open worldlines, i.e., from the change of intersection properties
upon opening a closed loop.

 In the present work, we have used a particular choice of the
 potential $\potential$ to impose Dirichlet boundary conditions for
 massless field fluctuations on Casimir surfaces. Generalizations to
 softer properties of the plates and to a finite-mass scalar field are
 straightforward. Away from the Casimir surfaces, our construction
 guarantees that the EMT components are finite.  Worldline numerics
 can then be used to calculate directly components of the
 energy-momentum tensor and check energy conditions.

 From a technical perspective, the use of a point-splitting method
  suggests employing \textit{common point} lines in the worldline
average instead of center-of-mass loops for quantum actions. In
contrast to the computation of effective actions, local operators do
not only involve powers but also derivatives of cumulants of
worldline variables. As any (propertime) lattice Monte Carlo
  calculation, our method comes with a statistical error controlled by
  the number $N$ of configurations (loops or lines) in the ensemble,
  as well as with a systematic error from the discretization of the
worldline propertime in terms of the number of points per loop
$n_{ppl}$. In addition, the approximation of derivatives with
difference quotients leads to further systematic errors. We have
demonstrated for several examples how to estimate and control
theses errors straightforwardly.

 On the physics side, we have used our new method to re-calculate
 several results for the canonical energy-momentum tensor induced by a
 minimally coupled massless Dirichlet scalar. The well-known violation
 of the null energy condition is reproduced as well. Furthermore, we
 have shown how the conformal energy-momentum tensor can be
 constructed from its canonical counterpart within our approach. The
 conformal EMT also violates the NEC. Our numerical algorithm proved
 to be highly efficient, with results within at worst $3\,\%$ of the
 analytic values at computation costs of about 1 CPU-day on standard
 desktop PCs. We believe that our new method is now sufficiently
 mature to study more involved Casimir geometries in order to check
 the status of more sophisticated energy conditions such as, e.g., the
 ANEC along the lines of \cite{Graham:2005cq} and beyond.

\appendix
\section{Evaluation of $\ImP\Geff$\label{sec:App1}}

In \secref{sec:WL_for_G} we expressed the Green's function $\Geff$ in the worldline formalism. Due to the functional structure of the components of the energy-momentum tensor in \EQref{eq:EMTG}, viz. \begin{math}T_{\cdot\cdot}\propto\int\mathrm{d}k\,f(k)\,\ImP\int\mathrm{d}s\,g(s)\,\er^{isk^2}\end{math}, the Wick rotations of $s$ and $k$ should not be performed independently but simultaneously. To do so, we pull the $k$ integral inside the argument of the imaginary part. We use the Minkowskian versions of \EQref{eq:freePI} and \EQref{eq:PIvev} as well as $E_k^2=m^2+k^2$ and thus find that only two different integrals ($\mathcal{I}_{1}$ and $\mathcal{I}_{2}$) occur in \EQref{eq:EMTG}:
\begin{align}\nonumber
 \mathcal{I}_{1} \overset{\hspace{1.75em}}{=} {} & 
 \ImP\left[i\int\limits_{0}^{\infty}\mathrm{d}s\,\frac{\er^{i\vec{\Delta}^2/s}}{\left(4\pi is\right)^{d/2}}\,\left\langle\er^{-i\int\limits_0^s\mathrm{d}\tau\,\sigma(\vec{x}(\tau))}-1\right\rangle\int\limits_0^\infty\mathrm{d}k\,k\sqrt{k^2+m^2}\er^{isk^2}\right]\\\nonumber
 \overset{\hspace{1.75em}}{=} {} & 
 \ImP\left[-\int\limits_{0}^{\infty}\mathrm{d}T\,\frac{\er^{-\vec{\Delta}^2/T}}{\left(4\pi T\right)^{d/2}}\,\left\langle\er^{-\int\limits_0^T\mathrm{d}\tau\,\sigma(\vec{x}(\tau))}-1\right\rangle\int\limits_0^\infty\mathrm{d}k_{E}\,k_{E}\sqrt{m^2-k_{E}^2}\er^{-Tk_{E}^2}\right]\\\nonumber
 \overset{\hspace{1.75em}}{=}{} & 
 -\int\limits_{0}^{\infty}\mathrm{d}T\,\frac{\er^{-\vec{\Delta}^2/T}}{\left(4\pi T\right)^{d/2}}\,\left\langle\er^{-\int\limits_0^T\mathrm{d}\tau\,\sigma(\vec{x}(\tau))}-1\right\rangle\ImP\left[\frac{i}{4T^{3/2}}\er^{-m^{2}T}2\Gamma\left[\frac{3}{2},-m^{2}T\right]\right]\\
 \overset{m\to0}{=} {} & 
 -\int\limits_{0}^{\infty}\mathrm{d}T\,\frac{\er^{-\vec{\Delta}^2/T}}{\left(4\pi T\right)^{d/2}}\,\left\langle\er^{-\int\limits_0^T\mathrm{d}\tau\,\sigma(\vec{x}(\tau))}-1\right\rangle\frac{1}{4T^{3/2}}\sqrt{\pi},
 \end{align}
 \begin{align}\nonumber
 \mathcal{I}_{2} \overset{\hspace{1.75em}}{=} {} & 
 \ImP\left[i\int\limits_{0}^{\infty}\mathrm{d}s\,\frac{\er^{i\vec{\Delta}^2/s}}{\left(4\pi is\right)^{d/2}}\,\left\langle\er^{-i\int\limits_0^s\mathrm{d}\tau\,\sigma(\vec{x}(\tau))}-1\right\rangle\int\limits_0^\infty\mathrm{d}k\,\frac{k}{\sqrt{k^2+m^2}}\er^{isk^2}\right]\\\nonumber
 \overset{\hspace{1.75em}}{=} {} & 
 \ImP\left[-\int\limits_{0}^{\infty}\mathrm{d}T\,\frac{\er^{-\vec{\Delta}^2/T}}{\left(4\pi T\right)^{d/2}}\,\left\langle\er^{-\int\limits_0^T\mathrm{d}\tau\,\sigma(\vec{x}(\tau))}-1\right\rangle\int\limits_0^\infty\mathrm{d}k_{E}\,\frac{k_E}{\sqrt{m^2-k_E^2}}\er^{-Tk_{E}^2}\right]\\\nonumber
 \overset{\hspace{1.75em}}{=} {} & 
 -\int\limits_{0}^{\infty}\mathrm{d}T\,\frac{\er^{-\vec{\Delta}^2/T}}{\left(4\pi T\right)^{d/2}}\,\left\langle\er^{-\int\limits_0^T\mathrm{d}\tau\,\sigma(\vec{x}(\tau))}-1\right\rangle\ImP\left[i\sqrt{\frac{\pi}{4T}}\er^{-m^2T}\left(\Erf\left[\sqrt{-m^{2}T}\right]-1\right)\right]\\
 \overset{m\to0}{=} {} & 
 \int\limits_{0}^{\infty}\mathrm{d}T\,\frac{\er^{-\vec{\Delta}^2/T}}{\left(4\pi T\right)^{d/2}}\,\left\langle\er^{-\int\limits_0^T\mathrm{d}\tau\,\sigma(\vec{x}(\tau))}-1\right\rangle\sqrt{\frac{\pi}{4T}}.
\end{align}
For both calculations we have performed the substitution $s=-iT$ and $k=-ik_{E}$ in the integration variables and the rotation of the integral contours in the second line. We assume $m>0$ and $T>0$ for convergence. In the last step we have considered $\fieldPhi$ to be massless. The massless limit is, however, not a necessity and
our calculations easily generalize to the case of finite $m$. It should be noted that the massless case is obtained as $m \to 0^{+}$, but can also be conceived on its own right by using $E_k = \sqrt{k^2}$, which guarantees the correct analytic continuation under the double Wick rotation.

\section{Propertime integration of derivatives of the $\Theta$ function\label{sec:App2}}

In \secref{sec:CompExpr} we compute compact expressions of EMT
components. This involves the interchange 
of several limits. Especially
the propertime integration of derivatives of the step function deserve
more details. These integrals can be solved by substituting
$q=\sqrt{T}\mathcal{M}-1$ and integrating by parts. Then the integrals
are of the following form:
\begin{align}
\begin{split}
\mathcal{A} = {} &
\int\limits_{-1}^{\infty}\mathrm{d}q\left(\frac{\mathcal{F}}{(q+1)^{\alpha}}\partial_{q}\Theta(q)+\frac{\mathcal{G}}{(q+1)^{\alpha-1}}\partial_{q}^2\Theta(q)\right)\\
= {} &
\lim_{r\to-1}\frac{\mathcal{G}}{(q+1)^{\alpha-1}}\delta(q)\bigg\vert_{r}^{\infty}+\int\limits_{-1}^{\infty}\mathrm{d}q\left(\frac{\mathcal{F}}{(q+1)^{\alpha}}+\frac{(\alpha-1)\mathcal{G}}{(q+1)^{\alpha}}\right)\delta(q).
\end{split}
\end{align}
Here $\mathcal{F}$ is a function of $\mathcal{M}$ and its second derivatives and $\mathcal{G}$ a function of $\mathcal{M}$ and its first derivatives with respect to either $\delta_{z}$ or $z_{cp}$. The exponent $\alpha$ is either $d+1$ or $d-1$. We identify $\partial_{q}\Theta(q)=\delta(q)$ since $q=0$ is within the domain of integration. This makes the remaining integral trivial to evaluate. The boundary term vanishes because $\delta(q)$ is zero everywhere except at $q=0$.

\begin{acknowledgments}
We would like to thank Noah Graham and Christian Schubert for
discussions.  We~have benefited from activities within the ESF
Research Network CASIMIR. MS was funded by the DFG Research Training
Group ``Quantum- and Gravitational Fields'' (GRK 1523). IH was partly
funded by the Conacyt grant 142561 and the SNI. HG acknowledges
support by the DFG under grant No.  Gi328/5-2.
\end{acknowledgments}

\bibliography{EMT_with_WL_numerics}

\end{document}